\documentclass[fleqn,usenatbib]{mnras}

\usepackage{newtxtext,newtxmath}

\usepackage[T1]{fontenc}

\usepackage{aecompl}
\usepackage{graphicx}	

\usepackage{xcolor}

\usepackage{subfig}
\usepackage{comment}
\usepackage{multirow}

\usepackage{mathrsfs}
\usepackage{orcidlink}

\newcommand{\quotes}[1]{``#1''}
\newcommand\code[1]{\textsc{\MakeLowercase{#1}}}

\def\be{\begin{equation}}
\def\ee{\end{equation}}

\def\msun{{\rm M}_{\odot}}

\def\msunpc2{\msun/{\rm pc}^{2}}

\def\gsim{\lower.5ex\hbox{\gtsima}} 
\def\lsim{\lower.5ex\hbox{\ltsima}} 
\def\gtsima{$\; \buildrel > \over \sim \;$} 
\def\ltsima{$\; \buildrel < \over \sim \;$} \def\gsim{\lower.5ex\hbox{\gtsima}} 
\def\lsim{\lower.5ex\hbox{\ltsima}} 
\def\simgt{\lower.5ex\hbox{\gtsima}} 
\def\simlt{\lower.5ex\hbox{\ltsima}} 

\def\nh2{n_{\rm H2}}

\definecolor{mkcolor}{HTML}{01abdf} 
\definecolor{apcolor}{HTML}{b3003b}
\definecolor{afcolor}{HTML}{01bdff}

\definecolor{blue-violet}{rgb}{0.54, 0.17, 0.89}
\definecolor{ao}{rgb}{0.0, 0.5, 0.0}
\definecolor{auburn}{rgb}{0.43, 0.21, 0.1}

\graphicspath{{plots/}}

\title[PINN to solve ISM chemistry]{Neural networks: solving the chemistry of the interstellar medium}

\author[Branca \& Pallottini]{
L. Branca\orcidlink{0000-0002-6064-1964}$^{1}$\thanks{\href{mailto:lorenzo.branca@sns.it}{lorenzo.branca@sns.it}},
A. Pallottini\orcidlink{0000-0002-7129-5761}$^{1}$
\\
$^{1}$ Scuola Normale Superiore, Piazza dei Cavalieri 7, I-56126 Pisa, Italy\\
}

\date{Accepted XXX. Received YYY; in original form ZZZ}

\pubyear{2022}

\begin{document}
\label{firstpage}
\pagerange{\pageref{firstpage}--\pageref{lastpage}}
\maketitle

\begin{abstract}
Non-equilibrium chemistry is a key process in the study of the InterStellar Medium (ISM), in particular the formation of molecular clouds and thus stars. However, computationally it is among the most difficult tasks to include in astrophysical simulations, because of the typically high (>40) number of reactions, the short evolutionary timescales (about $10^4$ times less than the ISM dynamical time) and the characteristic non-linearity and stiffness of the associated Ordinary Differential Equations system (ODEs).
In this proof of concept work, we show that Physics Informed Neural Networks (PINN) are a viable alternative to traditional ODE time integrators for stiff thermo-chemical systems, i.e. up to molecular hydrogen formation (9 species and 46 reactions).
Testing different chemical networks in a wide range of densities ($-2< \log n/{\rm cm}^{-3}< 3$) and temperatures ($1 < \log T/{\rm K}< 5$), we find that a basic architecture can give a comfortable convergence only for simplified chemical systems: to properly capture the sudden chemical and thermal variations a Deep Galerkin Method is needed.
Once trained ($\sim 10^3$ GPUhr), the PINN well reproduces the strong non-linear nature of the solutions (errors $\lsim 10\%$) and can give speed-ups up to a factor of $\sim 200$ with respect to traditional ODE solvers. Further, the latter have completion times that vary by about $\sim 30\%$ for different initial $n$ and $T$, while the PINN method gives negligible variations.
Both the speed-up and the potential improvement in load balancing imply that PINN-powered simulations are a very palatable way to solve complex chemical calculation in astrophysical and cosmological problems.
\end{abstract}


\section{Introduction}

Thermal and chemical evolution are crucial processes in astrophysical and cosmological environments. Gas cooling, heating, ionization, and photo-dissociation are vital drivers of the evolution of the interstellar (ISM) and intergalactic (IGM) medium.
Chemical processes are widely included in theoretical works and numerical simulations, playing key roles in determining the evolution in the early Universe history \citep{1998A&A...335..403G,glover:2008}, the IGM during the Epoch of the Reionization \citep{maio:2007,theuns:1998}, galaxy formation and evolution \citep{pallottini:2017_b,lupi:2019}, and Giant Molecular Cloud \citep[GMC][]{kim:2018,decataldo:2019}.

The typical chemical network involves hydrogen, helium, (possibly individual) metals, and molecules, by further coupling of all these various species with the radiation field (i.e. photoheating and photoionization), dynamics (i.e. shocks) and interaction with dust; moreover, going to progressively smaller scales, the chemical processes become more and more complex.
To follow a non-equilibrium chemical and thermal evolution in a numerical simulation is necessary an additional set of Ordinary Differential Equations  (ODEs) that describe the coupling of the various species.

Various numerical schemes and implementations have been developed to accomplish this task: \code{KROME} \citep{grassi:2014}, \code{XDELOAD} \citep{2005Ap&SS.299....1N}, \code{ASTROCheM} \citep{2013MNRAS.431..455K}, \code{ALCHEMIC} \citep{2010A&A...522A..42S}, \code{GRACKLE} \citep{smith:2017}, and \code{GGCHEMPY} \citep{2022RAA....22a5004G}.
The strategy/approximations adopted in a specific implementation can vary, however all schemes have to face similar key problems: i) the chemical ODE system is often stiff and ii) the typical time-scales are much shorter than the dynamical/hydrodynamic time , e.g. $\Delta t_{chem}\ll 10^{-4} \Delta t_{hydro}$.
Thus, adopting robust multi-step implicit schemes is needed in order to numerically solve the ODEs with  procedural methods \citep{byrne:19871}.
Such schemes often rely on matrix inversion, which -- at face value -- has a computational complexity that scales as $\mathcal{O}(N_{spec}^3)$, with $N_{spec}$ being the number of species in the network. The matrix associated with chemical ODEs is often sparse, which ameliorate the burden of the inversion \citep{grassi2021reducing}; nonetheless, systems experience a fast growth of the computational cost in including progressively more complex chemical network, ultimately making the CPU time spent on chemistry a relevant fraction of a hydrodynamical simulation.
Further, the precise computational time for the inversion is difficult to estimate given the ODEs, which thus can spoil the load-balancing of the typical astrophysical code.

Thus -- in order to try to overcome these limitations -- it is interesting to consider emulators as possible fast alternatives to a procedural resolution of chemical networks.
The usage of emulators in astrophysics, based on machine learning (ML) and deep learning techniques, has steadily increased its importance in recent years \citep{lecun2015deeplearning}.
However, so far ML applications in astrophysics are mainly limited to data driven inferences as parameters or classifications; for example \citet{ucci:2018} use decision trees to infer key ISM physical properties from emission line ratios, \citet[][]{chardin:2019} emulates radiative transfer calculation using Epoch of Reionization simulations as a training dataset, \citet{pregolovic:2022} infer astrophysical parameters from 21 cm light cone images by adopting recurrent neural networks, and \citet{dropulic:2021} shows how to predict stellar line-of-sight velocity from Gaia observations of the Milky Way (MW) by training on phase space mock data sets.
Currently, there are few attempts to alleviate the cost of computing chemistry by using auto-encoders: \citet{grassi2021reducing} tries to reduce the complexity of chemical ODE with high dimensionality by compression in a latent space of a smaller dimension; while \citet{grassi2021reducing} showcased the approach for isothermal models, its generalization seems non-trivial.
On the other hand, \citet{2021A&A...653A..76H} recently implement a fast emulator that includes temperature evolution; it works combining auto-encoders (to reduce the dimensionality) and emulators to follow the temperature and abundances evolution. 
All these solutions seem encouraging, promoting future usage of emulators for thermo-chemistry in hydrodynamic simulations. 
However, as these are supervised learning strategies, they require the creation of datasets containing the solution of the ODEs system using a traditional solver. 

An intriguing alternative consist in direct approximation of the solution for the chemical system with a neural network (NN), so that once the training is complete the chemical evolution can be directly computed at lower computational cost.
Solving differential equations with NN can be done adopting the Neural Ordinary Differential Equation (NODE) model \citep{2018arXiv180607366C}, that use Recurrent Neural Networks with an high number of layers to emulate the discretization of a differential operator.
A more general approach consist in using Physics Informed Neural Network (PINN) \citep{2019JCoPh.378..686R}. The PINN method is very flexible and  has been used to solve a large class of differential system associated with wide range of physical problems, as 2D acoustic wave equations \citep{2020arXiv200611894M}, turbulent fluid-dynamics \citep{hennigh2020nvidia} or full radiative transfer calculation \citep{2021JQSRT.27007705M}.

In this work we choose to focus on the PINN, in part because of its flexibility and in part since the mode seems simpler to extend with respect to the original implementation presented in \citet{2019JCoPh.378..686R}, by using the eXtended Physics Informed Neural Network \citep[X-PINN,][]{2022SJSC...44A3158H}.
While various implementation of the PINN frameworks have been developed \citep{ lu2020deepxde,Haghighat_2021,rackauckas2019diffeqfluxjl,hennigh2020nvidia}, applications in the astrophysical and cosmological context has yet to be explored, except in \citet{2022arXiv220502945C}, where PINN (in this works named Cosmological Informed neural Networks, CINN) are used to solve the background dynamics of the universe for four different models and then to perform statistical analyses to estimate the values of each model’s parameters with observational data.

To our knowledge, only \citet{Ji_2021} performs an in-depth study of stiff chemical systems via PINN, by showcasing the solution of standard ODE benchmarks, as the ROBER (3 non-linear ODEs) and POLLU problems (20 non-linear ODEs); however, the adopted reaction coefficients are constant, while for the typical ISM network they can vary by several orders of magnitude, mainly due to temperature\footnote{Formally, this is correct when only two body reactions are considered; for a general chemical network, further variation are present, e.g. photoionization reactions are proportional to the radiation field and, in cosmic ray induced reactions, their flux plays a similar role.}; such dependence is key for astrophysical problems and promotes the system from chemical to thermo-chemical, thus making its resolution more complex.

In this proof of concept work, we show that it is possible to train a PINN to emulate a complex and realistic chemical network that can be used to simulate the ISM.
Our aim is to showcase the PINN scheme in a non-trivial astrophysical context, by proposing an efficient and accurate alternative to the procedural solvers.

In Sec. \ref{sec:metodo} we introduce chemical networks and their usage in astrophysical context (Sec. \ref{sec:chimica_astro}), we present the general framework of the PINN method (Sec. \ref{sec:pinn_general}), and we benchmark PINN for simplified ODE systems (Sec. \ref{sec:benchmark}).
Then, in Sec. \ref{sec:PINNism}, we detail and implement the further modelling needed to adopt the PINN framework to treat realistic ISM chemical networks.
Sec. \ref{sec:results} presents the results of our models in terms of accuracy, efficiency, and in comparison with procedural solvers.
In Sec. \ref{conclusions} we give our conclusions.

\section{Methods overview}\label{sec:metodo}

In this Sec. we provide a summary of the thermal and chemical evolution of the ISM (Sec. \ref{sec:chimica_astro}), present the procedural tool used as reference to test our results, and introduce the usage of neural network solvers, in particular focusing on PINN algorithm (Sec. \ref{sec:pinn_general}), we benchmark the model for a chemical and thermo-chemical like ODEs system (Sec. \ref{sec:benchmark}).

\subsection{Chemistry of the Interstellar medium}\label{sec:chimica_astro}

\begin{figure}
    \centering
    \includegraphics[width=0.49\textwidth]{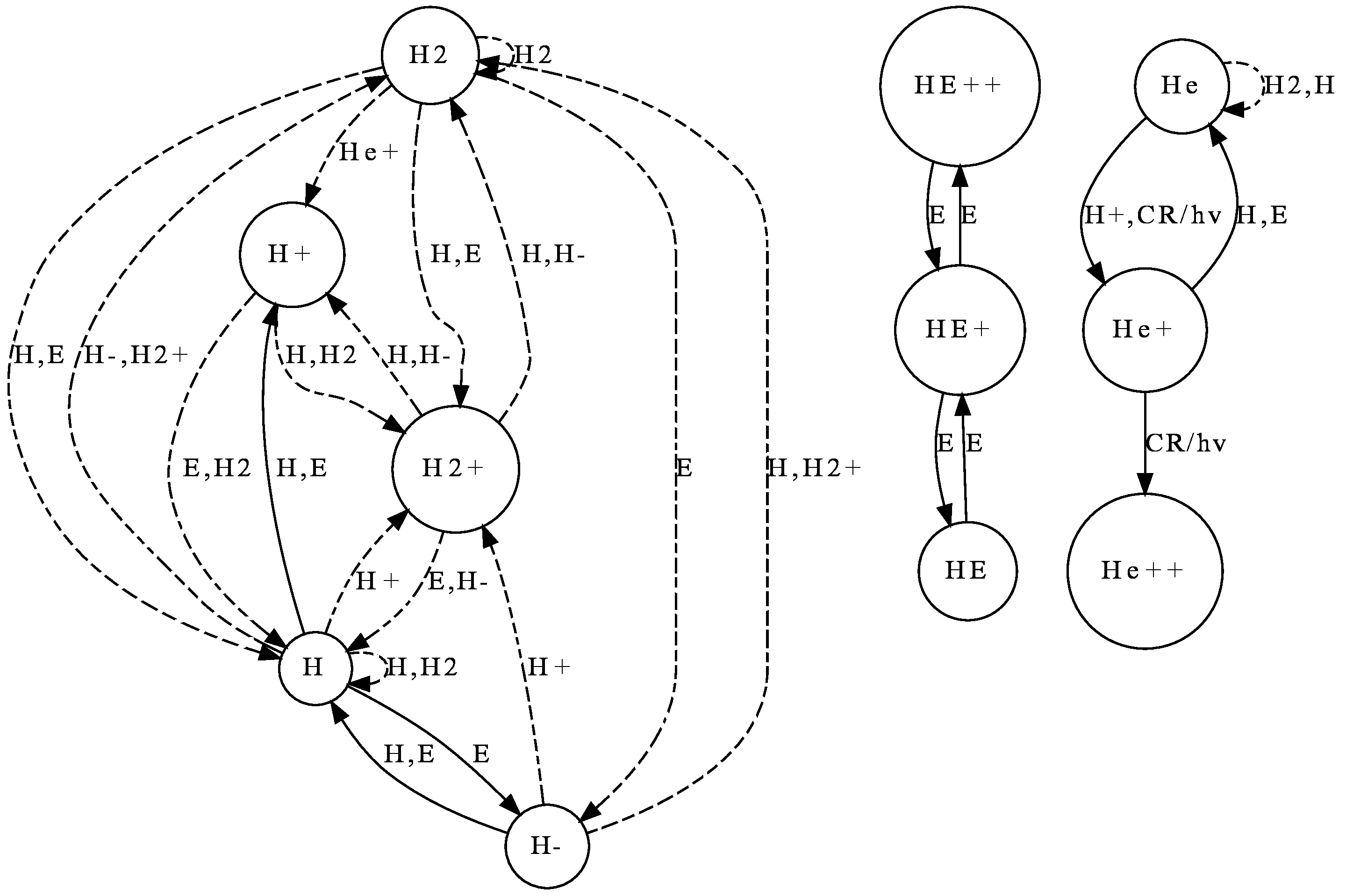}
    \caption{Sketch of the chemical networks adopted in this work. Circles represent the chemical species, while dashed and solid lines connected with arrows represents the two-body and photo-reactions (including cosmic rays). The dashed lines underline reactions involving molecular hydrogen.}
    \label{ism_net_scheme}
\end{figure}

Given a chemical network, following its evolution entails knowing the density of each species and the temperature at each time step.
This chemical and thermal evolution can be described by an ODEs system, with one equation for each  species involved and one for the temperature evolution. Each equation is given by a sum of terms that depends on the chemical reaction involved.

In general, reactions depends on the number density of each species ($\mathbf{n} = n_{1},\dots,n_{N_{spec}}$), the temperature ($T$) and the radiative flux (ionizing photons, photo-dissociation photons,  cosmic rays, ...)
The ODEs system for the chemical species can be written as \citep[e.g.][]{grassi:2014}
\begin{equation}
    \dot{n}_k = \sum_{j\in {\rm reaction}_{k}} \left(a_j \prod_{r \in {\rm reactant}_{j}} n_{r(j)}\right)\,,
    \label{react}
\end{equation}
where $n_k$ is the k-th species, and $a_j$ are the rate coefficients for all the reactions considered in the chemical network.

Considering only 2-body reactions and photo-reactions\footnote{Neglecting secondary higher order processes, also reaction involving cosmic rays can be approximated with the same formalism, see e.g. \citet{bovino:2016}.}, eq. \ref{react} can be re-written as:
\begin{equation}
    \dot{n}_k= A^{ij}_k n_i n_j+B_k^i n_i\,,
    \label{2body}
\end{equation}
where $A^{ij}_k=A^{ij}_k(T,\mathbf{n})$ are 2-body reaction coupling coefficients, $B^{i}_k=B^{i}_k(\mathbf{F})$ describe the photo-reactions rates, with $\mathbf{F}$ quantifying the photon and cosmic ray flux in various energy bins.

The evolution of the thermal state of the gas is accounted evolving the gas temperature, that depends on the heating and cooling process (chemical, radiative, ...):
\begin{equation}
    \label{temp_evolution}
    \dot{T}=\frac{(\gamma-1)}{k_b\sum_{i}n_i}(\Gamma-\Lambda)\,,
\end{equation}
where $k_b$ is the Boltzmann constant, $\gamma$ is the gas adiabatic index, $\Gamma=\Gamma(T,\mathbf{n},\mathbf{F})$ and $\Lambda=\Lambda(T,\mathbf{n},\mathbf{F})$ are the heating and cooling functions, respectively.

In this work we focus on a ISM chemical network originally presented in \citet{bovino:2016} and that has been used for studies on molecular cloud scales \citep[][]{decataldo:2019,decataldo:2020} and the on evolution of high-redshift galaxies \citep[][]{pallottini:2017_b,pallottini:2019,pallottini:2022}.
The network has $N_{spec}=9$ species: $\mathrm{e}^-$, $\mathrm{H}^-$, $\rm H$, $\mathrm{H}^+$, He, $\mathrm{He}^+$, $\mathrm{He}^{++}$, $\mathrm{H}_2$, and $\mathrm{H}_2^+$.
Following \citet{bovino:2016}, the evolution of the species is regulated by 46 reactions (for a schematic view, see Fig. \ref{ism_net_scheme}), involving dust processes, i.e. H$_2$ formation on dust grains \citep{1975ApJ...197..575J}, photo-chemistry, and cosmic rays ionization. In particular, the rates are taken from \citet{bovino:2016}: reactions 1 to 31, 53, 54, and from 58 to 61 in their tables B.1 and B.2, photo-reactions P1 to P9 in their table 2.

For the temperature evolution (eq. \ref{temp_evolution}) we account for the following processes: photoelectric heating from dust \citep{1994ApJ...427..822B}, cosmic rays heating \citep{1992ApJS...78..341C}, photo heating, heating/cooling due to exothermic/endothermic reactions, metal line cooling \citep{2013ApJ...765...89S}, Compton cooling from the CMB, molecular $\mathrm{H}_2$ cooling \citep{glover:2008}, and atomic cooling \citep{1992ApJS...78..341C}.
For simplicity, in this work, we adopt a constant solar value for the metallicity ($Z=Z_{\odot}$, \citet{2009ARA&A..47..481A}) and dust to gas ratio ($f_d=0.3$, \citet{hirashita:2002}).

Recall that, the two body reactions ($A^{ij}_k$, eq. \ref{2body}) depends only on density $\mathbf{n}$ and the temperature $T$, however the coefficients of ($B^i_k$, eq. \ref{2body}), and the heating and cooling terms ($\Gamma$ and $\Lambda$, eq. \ref{temp_evolution}) additionally depends on the flux $\mathbf{F}$.

For simplicity, in this work we consider a Spectral Energy Distribution (SED) of UV/Xray background from \citet{2012ApJ...746..125H} at redshift $z=0$ and adopt MW like cosmic ray flux with rate $\zeta_{cr}=3\times 10^{-17}\mathrm{s}^{-1}$.
This SED is not completely appropriate for the typical ISM conditions, \citep[e.g. see][for the MW]{1978ApJS...36..595D}, however we adopt it so that all photo-ionizations ($\rm H+\gamma_{{\rm h}\nu > 13.6 eV}\rightarrow H^{+}+e$, ...) in our chemical network are active, i.e. this choice allow us to robustly test all the reactions in the model.

Various implementations/schemes can be adopted to solve a chemical network. As a reference for this work, we adopt the flexible code \code{KROME}\footnote{\url{https://bitbucket.org/tgrassi/krome/src/master/}} \citep{grassi:2014}, which is a framework that -- given an input chemical network -- generates the code to solve the associated ODE system.
To solve the system \code{KROME} use \code{LSODES}, which is included in \code{ODEPACK} \citep{odepack}. \code{LSODES} is an implicit, robust, multistep, iterative high order solver (5 by default) that can take advantage of the sparsity of the Jacobian matrix of the ODEs.
The default \code{KROME} relative and absolute tolerances are fixed at $10^{-4}$ and $10^{-20}$ respectively.

In this work, we adopt \code{KROME} i) to build the ODEs structure (eq.s \ref{2body} and \ref{temp_evolution}) for our PINN scheme (Sec. \ref{sec:PINNism}) and ii) to test our results during the validation phase (Sec. \ref{sec:results}).

\subsection{Physics Informed Neural Network}\label{sec:pinn_general}

\begin{figure}
    \centering
    \includegraphics[width=0.49\textwidth]{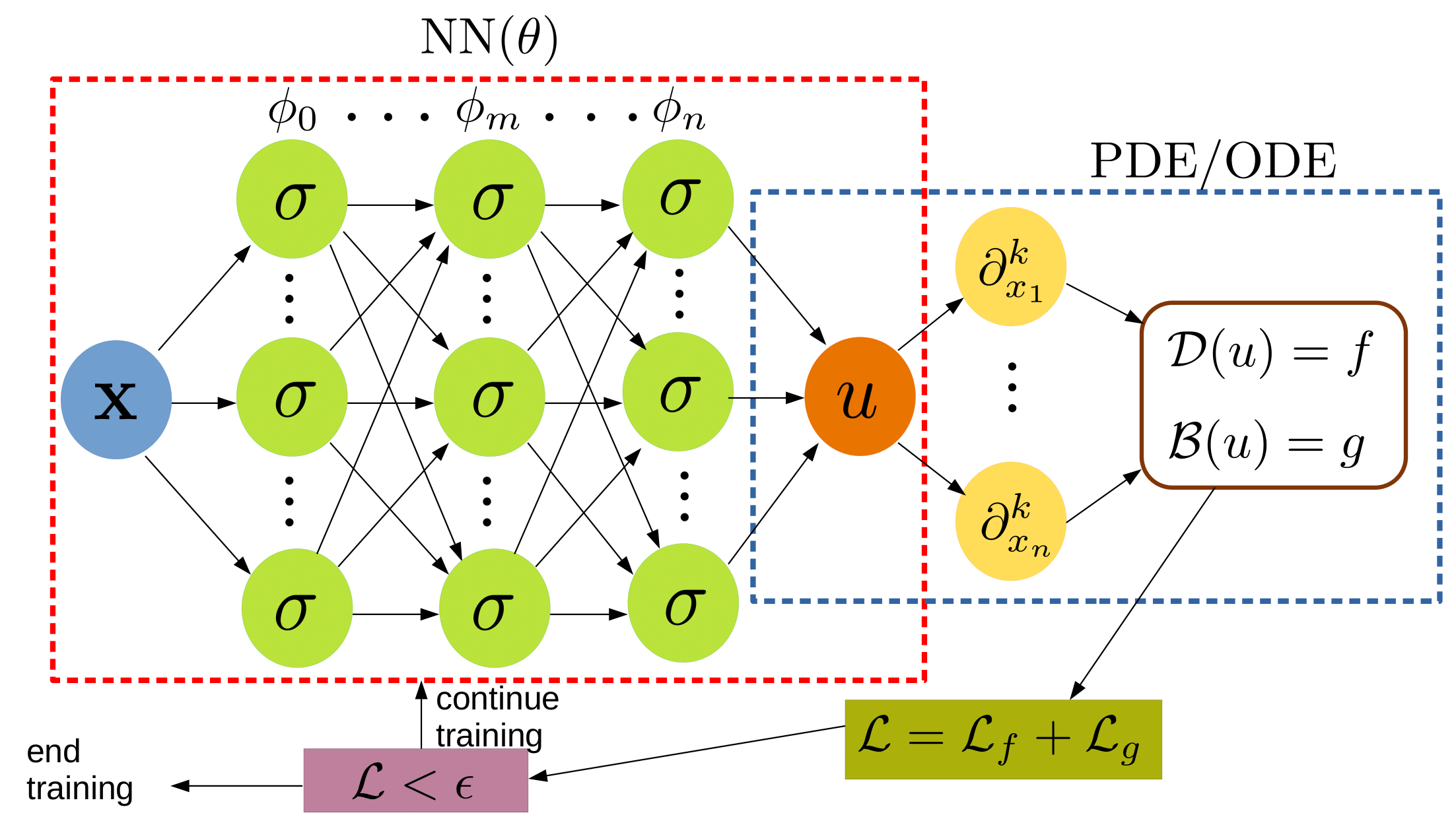}
    \caption{General scheme for a feed-forward Neural Network (NN) architecture. The aim is to resolve the PDE/ODE system (eq. \ref{eq:pde_system}) of the set of variables $\textbf{x}$.
    This variables are used as the input for the NN: $\textbf{x}$ passes through the Physics Informed NN (PINN) layers defined by the activation functions $\sigma$ (eq. \ref{eq:network_layer}) and the set of parameters $\theta$ (eq. \ref{eq:network_parameters}).
    The NN returns the emulated solution $u$, which is tested against the PDE/ODE system (eq. \ref{eq:pde_system}) by evaluating the residuals via the loss function (eq. \ref{eq:loss_function}).
    Parameters $\theta$ are updated and the process is repeated until convergence is reached (eq. \ref{eq:best}).
    \label{PINN_scheme}
    }
\end{figure}

In general, we can write a set of partial differential equations (PDE)/ODE\footnote{While in the present we are focusing on ODE systems associated with a chemical network, the PINN scheme can also be applied to PDE.} in the form

\begin{subequations}\label{eq:pde_system}
\begin{align}\label{first}
    \mathcal{D}(u(\mathbf{x}))&=f(\mathbf{x}), \; \forall\mathbf{x}\in \Omega, \\ 
    \mathcal{B}(u(\mathbf{x}))&=g(\mathbf{x}), \;  \forall \mathbf{x}\in \partial\Omega\,,
\end{align}
\end{subequations}
where $\mathbf{x}$ is the set of independent variables, $\mathcal{D}$ is the differential operator of the PDE/ODE, $\mathcal{B}$ is a constrain operator -- i.e. it represents the boundary/initial conditions (BC/IC) -- and $u(\mathbf{x})$ is the solution of the PDE/ODE system.

Our aim is to approximate the solution of the system with a neural network \citep[see][for a review analyzing the theoretical and practical aspects]{2017arXiv170207800W}. In principle, this is possible because of the universal approximation theorem \citep{citeulike:3561150}, since  multi-layer feed-forward neural networks are capable of approximating any Borel measurable function \citep{HornikEtAl89}.
Namely, it is formally possible to replace the PDE/ODE solution $u(\mathbf{x})$ with the output of the neural network $u_{net}(\mathbf{x},\theta)$, which can be written as
\begin{equation}
        u_{net}(\mathbf{x},\theta)=\mathbf{W}_n(\phi_{n-1}\circ\phi_{n-2}\circ ... \circ\phi_1\phi_0)(\mathbf{x})+\mathbf{b}_n\,,
        \label{eq:network_layer_scheme}
\end{equation}
i.e. the composition of the action of successive layers $\phi_i$
\begin{equation}\label{eq:network_layer}    
\phi_i(\mathbf{x})=\sigma(\mathbf{W}_i\mathbf{x}+\mathbf{b}_i)\,,
\end{equation}
where $\sigma$ is the activation function, $\mathbf{W}_i$ is the matrix of the weights and $\mathbf{b}_i$ is the bias vector. It is convenient to cluster $\mathbf{W}_i$ and $\mathbf{b}_i$ as
\begin{equation}\label{eq:network_parameters}
\theta = \{\mathbf{W}, \mathbf{b}\}\,,
\end{equation}
i.e. the set of the parameters to be trained.
For the PINN method, the necessary condition is that $u_{net}$ is derivable at least $p$ times -- i.e. $u_{net(\mathbf{x},\theta)}\in \mathcal{C}^p$ -- with $p$ being the maximum derivative order for the operator $\mathcal{D}$ in eq. \ref{first}.

Using the auto-differentiation \citep{baydin2015automatic} it is possible to define the partial (or total) derivative $\partial_i u_{net}(x_i,\theta)$ by selecting a proper activation function $\sigma$, e.g. a typically a sigmoid or hyperbolic tangent. The activation functions are often chosen among smooth ($\mathcal{C}^\infty$) analytical functions, so that the derivatives calculated by auto-differentiation are correct at machine precision, which is important for the numerical stability in the evaluation of the PDE/ODE terms.

The core of the PINN scheme consist in evaluating the residual in the space of the neural network parameters $\theta$ (eq. \ref{first}). The aim is to optimize the \textit{loss function} $\mathcal{L}$, that can be defined as follows:
\begin{subequations}\label{eq:loss_function}
\begin{align}
  \mathcal{L}_f(\theta)    &= d_{\Omega}(\mathcal{D}(u_{net}(\mathbf{x},\theta))-f(\mathbf{x}, \theta)),\label{eq:loss_f} \\
  \mathcal{L}_g(\theta)    &= d^{'}_{\partial\Omega}(\mathcal{B}(u_{net}(\mathbf{x},\theta))-g(\mathbf{x},\theta)),\label{eq:loss_g} \\
  \mathcal{L}_{tot}(\theta)&= \mathcal{L}_f(\theta)+\mathcal{L}_g(\theta)
\end{align}
\end{subequations}
where $d_{\Omega}$ and $d^{'}_{\partial\Omega}$ are positive metrics that can be used to evaluate the distance to the exact solution $u$ and IC/BC, e.g. the $L_2$ norm ($\|\cdot\|_2^2$). Including the ODE/PDEs residuals in the loss function is the reason why the algorithm is called \textit{physics informed}.

In this context the residuals takes the place of what labels would be in a context of supervised learning, making the algorithm completely data independent (i.e. in this case, the  solutions obtained with a numerical solver). {Note that the metrics in eq.s \ref{eq:loss_function} can be very general, e.g. for a ODE/PDE system each equation (and associated IC/BC term) can be weighted individually (see later Sec. \ref{sec:loss_ISM}, in particular eq. \ref{eq:loss_composition}).

Once the loss function is defined, an optimization algorithm is applied on the parameters set $\theta$.  Various solution have been proposed for different problem, the most classical being the Stochastic Gradient Descent (SGD), i.e. a Gradient Descent (GD) algorithm applied on a random subset of the parameter space \citep{Robbins&Monro:1951}.
The general idea beyond the GD method is to evaluate the gradient in a set of points of the domain and then  mediate on them to find the optimal direction to minimize the loss function. However, when the domain is to huge (in terms of memory or in computational cost for the loss function gradient evaluation) or when the algorithm is distributed on more than one device (i.g. multi GPUs), the  SGD is preferable: the method is equivalent to GD but on a subset of points (usually named \textit{batch}). The parameters are update as follows:
\begin{equation}
    \theta_{i+1}= \theta_i-\eta \nabla\mathcal{L}(\theta_i)\,,
    \label{SGD}
\end{equation}
where $\eta$ is a scalar called learning rate.
In this work we use the more sophisticated SGD variant with adaptive learning and gradient momentum rate ADAM \citep{kingma2014adam} where the parameters update is defined as:
\begin{equation}
    \label{eq:adam}
     \theta_{i+1}= \theta_i -\eta\frac{m}{\sqrt{v}+\epsilon}\,,
\end{equation}
where $m$ and $v$ are the first and second moments of the gradient, and $\epsilon$ is the smoothing term. 
The procedure can be considered concluded if the algorithm finds a set of parameters named $\theta^*$ such that for a given positive scalar $\delta$ the following condition is met:
\begin{subequations}\label{eq:best} 
\begin{equation}
    \mathcal{L}_{tot}(\theta^*)<\delta\,,
\end{equation}
which implies that:    
\begin{equation}
    u_{net}(\mathbf{x},\theta^*)\simeq u(\mathbf{x})
\end{equation}
\end{subequations}
where $\delta$ is the required tolerance, that could be fixed a priori or determined on the fly, i.e. if the loss function is not decreasing for a large number of epochs.

In Fig. \ref{PINN_scheme} we show the sketch of the feed-forward architecture of the PINN that is used in this work. To summarize, the proposed method follows these steps: 
\begin{itemize}
    \item[0)] We define a sampling of the input parameter space $\mathbf{X}$ and we randomly initialize the PINN parameters $\theta$.
    \item[1)] Given $\theta$, we evaluate the solution of the PDE/ODEs system (eq.s \ref{eq:pde_system}) on the input space $\mathbf{X}$ using the PINN, i.e. $u=\mathrm{NN}_{\theta}(\mathbf{X})$
    \item[2)] We compute the partial derivatives that appear in the PDE/ODEs via automatic differentiation, thus we evaluate the residuals in the loss function defined in eq. \ref{eq:loss_function}.
    \item[3)] We update the values of parameters $\theta$ with the optimizer in eq. \ref{eq:adam}, repeating 1)-3) until the convergence is reached (eq. \ref{eq:best}).
\end{itemize}

The architecture is shown in Fig. \ref{PINN_scheme} is the first introduced in the literature, and therefore the simplest. However, in recent years, more complex schemes have been drawn, such as: Fourier Network (FN,  \citet{2020arXiv200610739T}) and its variations Modified Fourier Network (MFN, \citet{2021SJSC...43A3055W}), Highway Fourier Network, inspired by Highway Network first introduced in \citep{2015arXiv150500387S} that are designed explicitly to take in account high frequency variations of the solution. Other variants are: SInusoidal REpresentation NEtwork (SIREN \citet{2020arXiv200609661S}), mainly designed for periodic-like solutions, MeshfreeFlowNet \citet{2020arXiv200501463J}), designed for super-resolution tasks and the Deep Galerkin Method (DGM \citep{2018JCoPh.375.1339S}) which is inspired by Long Short Term Memory (LSTM) architecture but optimizes for PDEs computing.

\subsection{Model benchmark}\label{sec:benchmark}

\begin{figure*}
    \centering
    \includegraphics[width=\textwidth]{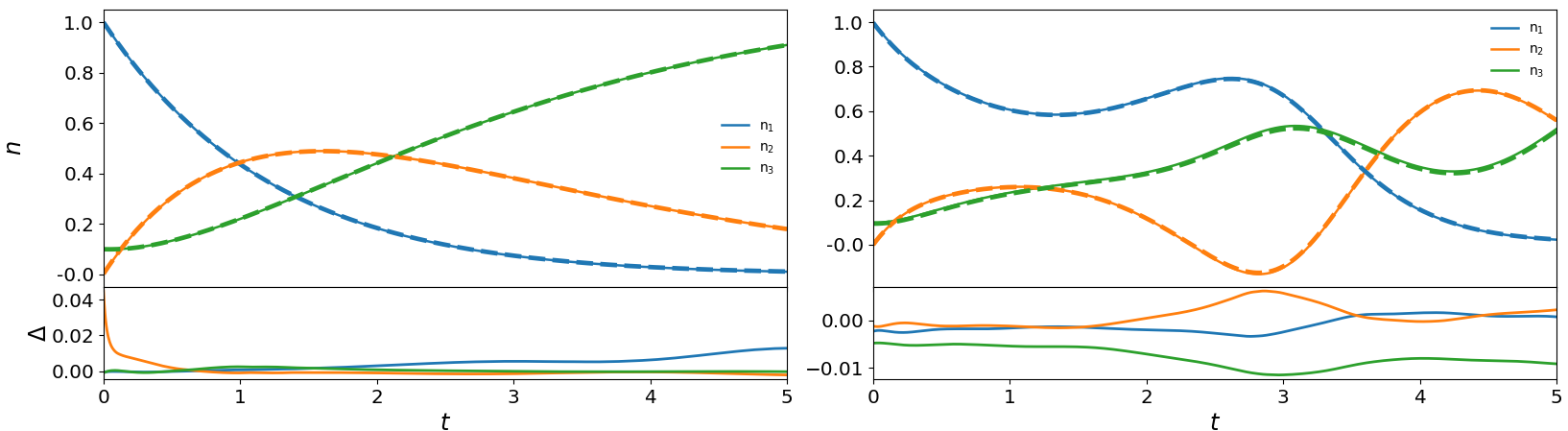}
    \caption{Benchmarks between the PINN and a procedural solver (Backward Differentiation Formula, BDF) for two simple ODEs systems with coefficients matrices given in eq. \ref{eq:matrix_shape}.
    Left (right) panel show the system without (with) explicit time dependence that is given in eq. \ref{eq:benchmark:const} (eq. \ref{eq:benchmark:tdep})).
    For each system, in the upper panel we show the comparison between our model (solid line) and BDF solver (dashed line), while the lower panel shows the absolute errors between the two methods (eq. \ref{eq:abs_err}). 
    \label{fig:benchmark}
    }    
\end{figure*}

Before trying to solve the ISM chemistry via the PINN method, we validate the algorithm on a simplified and easily reproducible problem.

For this benchmark, we select an ODE system similar in shape to eq. \ref{2body}. To have a formal description of the system, we rearrange the coefficient tensor as follows:
\begin{equation}
    \label{eq:matrix_formal}
    M^j_k = A^{ij}_kn_i +B^i_k\,,
\end{equation}
so that we write the ODE system in a compact (and standard) way 
\begin{equation}
    \dot{n}_k=M^j_k{n_j}\,,
    \label{eq:lin_ode}
\end{equation}
where $\mathbf{n}=(\mathrm{n}_1,\mathrm{n}_2,\mathrm{n}_3)$ is a 3-dimensional vector of \textit{fake species}, and $M$ is the rate coefficients matrix.

To give a proof of the performance of the model, we adopt the following shape for the matrix of coefficients
\begin{subequations}\label{eq:benchmark_ode}
\begin{equation}
\label{eq:matrix_shape}
M=\Bigg(\begin{array}{*{20}{c}} -(k_1+k_3n_3) & 0 & k_3n_1\cr k_1++2k_3n_3 & -k_2 & 2k_3n_1 \cr -k_3n_0 & k_1 & -k_3n_0 \end{array}\Bigg)
\end{equation}
and consider two different cases. For case \textbf{1)} we select rate coefficients as
\begin{equation}\label{eq:benchmark:const}
\mathbf{k}_{cons}=[0.8, 0.5, 0.2]\,,
\end{equation}
i.e. the coupling matrix depend only on density $M=M(\mathbf{n})$, which mimics a temperature-independent system. For \textbf{2)} we adopt
\begin{equation}\label{eq:benchmark:tdep}
\mathbf{k}_{t-dep}=[0.8-\sin(t), 0.5 +\cos(t), 0.2 + \sin(2t)]\,,
\end{equation}
\end{subequations}
so that $M=M(\mathbf{n},t)$, which mimics a temperature-dependent system. 
The initial conditions of the fake species for both models are $\mathrm{n}_1^{in}=1$, $\mathrm{n}_2^{in}=0.001$ and $\mathrm{n}_3^{in}=0.1$. For both models we consider the \textit{fake time} interval [0,5).

A key obstacle in solving a chemical (or thermo-chemical) evolution of a system is represented by stiffness. The ratio $\mathcal{S}$ can be used to quantify the stiffness of the system
\begin{equation}
    \label{eq:stifness}
    \mathcal{S}=\frac{|\mathrm{Re}(\lambda)|_{max}}{|\mathrm{Re}(\lambda)|_{min}}\,,
\end{equation}
where $\mathrm{Re}(\lambda)_{max/min}$ are the largest/smallest value of the real part of eigenvalues of the matrix. For both our test cases the coefficients change as the system evolves, thus we compute the mean stiffness ratio in the \textit{fake time} interval, obtaining $\langle\mathcal{S}(M(\mathbf{n}))\rangle\simeq 2.8\times 10^3$ and $\langle\mathcal{S}(M(\mathbf{n},t))\rangle\simeq 2.9\times 10^5$ for \textbf{1)} and \textbf{2)}, respectively. The difference in the stiffness ratio reflects the increase in complexity for a variable temperature system.

Thus, for the model with no explicit time dependence, we choose a neural network architecture with 6 layers of 64 neurons each. We adopt $2^{10}\simeq 10^3$ training points and -- to assure convergence of the loss function -- we run the training for $1.5\times 10^4$ epochs, which takes $\sim 0.2 \mathrm{CPUhr}$ on a INTEL i7-9700 CPU using the ADAM algorithm introduced in Sec. \ref{sec:pinn_general}.
Instead, for the model with explicit time dependence, the setup consists of 6 layers of 128 neurons each, $2^{13}\simeq 8\times 10^3$ training points, and $3\times 10^4$ epochs ($\sim 1.1\mathrm{CPUhr}$), reflecting the increased complexity.
For this benchmark we have selected a feed-forward architecture, that is the simpler setup which can achieve satisfactory results.
In order to compare our model with a procedural solver, we solve the systems with the Backward Differentiation Formula (BDF) method adopting the \code{ScyPy} implementation \citep{scipy2019}.

In Fig. \ref{fig:benchmark} we show the benchmark validation. Qualitatively, the reconstruction of the trained models (dashed line) seem good in both cases. Quantitatively, we can define the errors as\footnote{For the simple benchmark, it is convenient to avoid using relative error, since all the \textit{fake species} densities are order unit.}
\begin{equation}
    \label{eq:abs_err}
    \Delta = y_{BDF}-y_{NN}\,.
\end{equation}
The evolution of the error is show in the bottom panels of Fig. \ref{fig:benchmark}. For the model without/with explicit time dependence the mean errors are $\langle\Delta\rangle\simeq 10^{-3}$ and $2\times 10^{-3}$ respectively, which we consider a satisfactory result for our benchmark.
Interestingly, Fig. \ref{fig:benchmark} shows no evidence of a growth of the error during the time evolution, as it might be expected from a regular procedural method. Such behaviour is intrinsic of the PINN method, as the training algorithm aims to minimizing the residuals on the entire time domain simultaneously.
 
\section{Neural networks for ISM chemistry}\label{sec:PINNism}

While the benchmark shows good potential in applying of the PINN method for ISM chemistry, multiple aspects need additional considerations. Specifically, the initial conditions must be generalized (Sec. \ref{sec:initial_conditions}), the chemical network requires further consideration (Sec. \ref{sec:chemical_networks_ISM}), the loss function requires modifications (Sec. \ref{sec:loss_ISM}), the neural network must be improved, and the training process needs more careful attention (Sec. \ref{sec:network_struct_ISM}).

\subsection{ODE structure of the chemical networks}\label{sec:chemical_networks_ISM}

For the selected chemical network (Sec. \ref{sec:chimica_astro}), the associated ODEs system has non-trivial dependence of the coefficients on the temperature.
This is determined by the elements of the interaction matrix $M^i_k$ (defined in eq. \ref{eq:matrix_formal}): an example is shown in Fig. \ref{fig:matrix_elements}, where we plot a subset of the matrix element as a function of temperature.
In the allowed temperature range, the selected $M^i_k$ can vary by more than about 10 order of magnitude.
The stiffness of the system is higher with respect to the benchmark (eq. \ref{eq:benchmark_ode}); for instance, for normal ISM densities (i.e. the IC used in Sec. \ref{sec:initial_conditions}), the stiffness can reach values of $\mathcal{S}\sim 10^{16}$ for $T<2.5\times 10^4\mathrm{K}$ and $\mathcal{S}\sim 10^{4}$ for $T>2.5\times 10^4\mathrm{K}$.
This is a hint that i) the network architecture must be expanded ii) the training will be more expensive, also because of the generalized initial conditions.

Given these expectations and in order to explore different strategies for both the neural network and training, for the chemical system we consider both a \textit{molecular} network (introduced in Sec. \ref{sec:chimica_astro}) and a reduced \textit{atomic} network.
The \textit{atomic} network is a simplified version of the \textit{molecular} one, namely, it does not include the chemistry of molecular hydrogen and its cooling, simplifying both the reaction network and the temperature evolution. With respect to \textit{molecular}, in \textit{atomic} the number of species decreases from 9 to 7 and the number of reactions decreases from 46 to 24.

\begin{figure}
    \centering
    \includegraphics[width=0.49\textwidth]{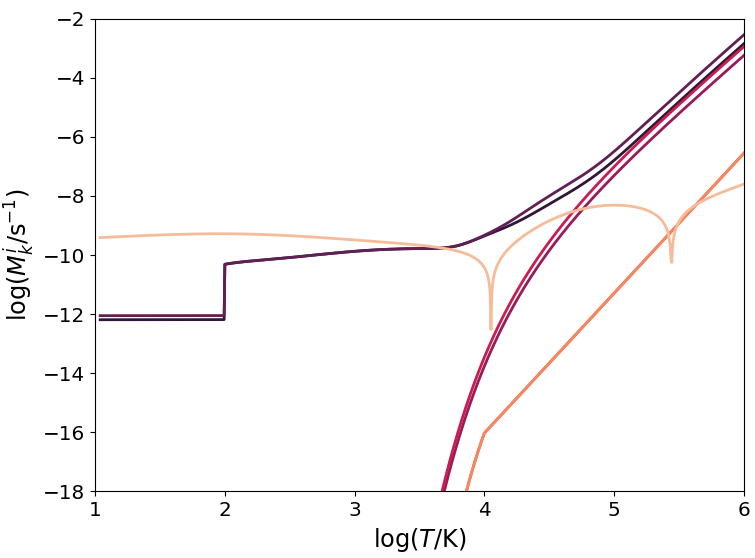}
    \caption{Example of temperature (T) dependence of matrix elements ($M^i_k$, definition in eq. \ref{eq:matrix_formal}.) of the adopted chemical network. We select 6 representative $M^i_k$ out of the 62 total non-null element (see Sec. \ref{sec:chimica_astro})}
    \label{fig:matrix_elements}
\end{figure}

\subsection{Setting up the initial conditions}\label{sec:initial_conditions}

The PINN model adopted in the benchmark (Sec. \ref{sec:benchmark}) is solved only with a specific set of fixed initial conditions: this is a major limitation.
Adopting the same approach for ISM chemistry would considerably limit the applicability of the model, since training is expensive and would be needed for each different thermodynamic configuration.
We can overcome this limit by generalizing the model as follows.

Recalling eq. \ref{eq:pde_system}, in the case of ODEs the operator it follows that $\mathcal{B}(\mathbf{x})=(x_1(t_0), x_2(t_0),...,x_k(t_0))$. Formally, to generalize for arbitrary initial conditions, we can promote the initial values from a scalar to a function $x_i(t_0)\rightarrow f_i(t_0)$ that map the \textbf{ICs} in a desire range, e.g. for temperature, $f_T(t_0)=[20, 10^6]\mathrm{K}$.
The procedure has been proposed by \citet{2020arXiv200614372F} and greatly increases the dimensionality of the problem; however, as shown by \citet{2020arXiv200701138M}, the PINN does not suffer too much from the \textit{curse of dimensionality}, e.g. can be trained in a 100-dimensional space in case of heat equation.
Summarizing, we can use this strategy to vary both the initial thermodynamic state ($T_{in}$, $n_{in}$) and the fractions of each species, $C_k$. We expect the procedure to increase the training time, however note that -- once the training is completed -- the computational time for predictions is mostly unaltered, since the latter depends only on the network architecture, i.e. the number of neurons and links.

Once trained our model maps the input space (time and initial conditions) to the evolution of the desired quantities:
\begin{equation}
    (t, \mathbf{IC})\rightarrow \mathrm{PINN}(t, \mathbf{IC})=(\mathbf{n}(t,\mathbf{IC}),T(t,\mathbf{IC}))\,.
    \label{eq:map}
\end{equation}

Stating eq. \ref{eq:map} differently, knowing the initial densities and temperature at the initial time $t=t_{in}$, the PINN return the evolved quantities at $t=t_{out}$:
\begin{align}
\left[n_k(t=t_{in}), T(t=t_{in})\right]   \rightarrow & \mathrm{PINN}(t_{out}, n_k(t=t_{in}), T(t=t_{in}))= \nonumber\\
                                         & = \left[n_k(t=t_{out}), T(t=t_{out})\right]
    \label{eq:map_ISM}
\end{align}
with $k$ indexing the species.

Note that the adopted generalization for the IC (eq. \ref{eq:map}) is not directly applicable to arbitrary PDE systems, which require more sophisticated and problem dependent techniques \citep[i.e.][]{inproceedings}.

Therefore, in this paper we distinguish 2 type of model: \textit{single models}, that have fixed initial condition, and \textit{general model}. Although single models are subclasses of the general models, in the spirit of proof of concept it is convenient to keep the cases separate, since different strategies are adopted to achieve convergence in the training.

For \textit{single models}, we adopt $T=10^{3}\mathrm{K}$ as the initial temperature for both networks (\textit{atomic/molecular}), while the total density ($n_{tot} \equiv \sum_{k}n_k$) and individual abundances ($C_k \equiv n_k/\sum_{k}n_k$) are set as follows.
\textit{atomic}: $n_{tot}=90.4\mathrm{cm}^{-3}$, $C_{\mathrm{H}^-}=0.0015$, $C_{\mathrm{H}}=0.69$, $C_{\mathrm{He}}=0.288$, $C_{\mathrm{H}^+}=0.0069$, $C_{\mathrm{He}^+}=0.0029$ and $C_{\mathrm{He}^{++}}=0.0008$;
\textit{molecular}: $n_{tot} = 100\mathrm{cm}^{-3}$, $C_{\mathrm{H}}=0.6241$, $C_{\mathrm{H}^-}=0.001$, $C_{\mathrm{H}_2}=0.104$, $C_{\mathrm{He}}=0.26$, $C_{\mathrm{H}^+}=0.0062$, $C_{\mathrm{H}_2^+}=0.001$, $C_{\mathrm{He}^+}=0.0026$ and $C_{\mathrm{He}^{++}}=0.0007$.
In both cases, $C_{\mathrm{e}^-}$ is set to ensure charge neutrality.

For \textit{general models}, the initial values for the temperature, total density, and individual abundances span the following ranges:
\begin{subequations}
\begin{align}
20 \leq & \frac{T}{\mathrm{K}} \leq 10^6\\
10^{-2} \leq & \frac{n_{tot}}{\mathrm{cm}^{-3}}\leq 10^{3}\\
10^{-6} \lsim & C_k \lsim 1\,,
\end{align}
\end{subequations}
where $C_k$ are chosen such that global charge neutrality is respected and the total hydrogen (helium) fraction is $X\simeq 70\%$ ($Y\simeq 30\%$).

\subsection{Loss function definition}\label{sec:loss_ISM}

A crucial aspect in any ML approach is the design of the loss function. Directly evaluating the mean square errors using the metric in eq. \ref{eq:loss_function} cannot capture the fine structure of the underlying solution. This is mainly driven by the large dynamical range spanned by the density of different species.
For instance, for typical ISM conditions H$^{-}$ is usually about 8 order of magnitude lower than H: using a simple (e.g. uniformly weighted) loss function would ignore the variation of H$^{-}$.
Ideally, to maximize the convergence efficiency, all components of the loss function must be of the same order of magnitude. However, unlike in a supervised learning problem, the code does not know in advance the abundances (and temperature) during the evolution.

With this in mind, we have adopted the following strategy to model the loss function. The first step consists in considering the evolution of the logarithm of the abundances (eq. \ref{2body}) and temperature (eq. \ref{temp_evolution}); this \quotes{feature normalization} is a relatively standard technique for ML.
Moreover, we solve the ODEs system in a logarithmic time scale; with this precaution we can better capture the sudden ($t\lsim 10^2 \, \rm yr$) large (more than one order of magnitude) variations that species can experience for some initial conditions.

To summarize, in linear space the ODE in eq. \ref{2body} has the following residuals:
\begin{equation}\label{def_residual_linear}
\mathcal{R}_{n_k} = \dot{n}_k - (A_k^{ij}n_j+B_k^i)n_i\,.
\end{equation}
We perform the change of variables:
\begin{subequations}\label{logsystem}
\begin{align}
    y_k       &= \log(n_k/{\rm cm}^{-3}) \\
    \tilde{T} &= \log(T/{\rm eV}) \\
    \tau      &= \log(t /{\rm s})  \,. \label{eq:sub:timelog}
\end{align}
\end{subequations}
Thus eq. \ref{2body} can be rewritten as
\begin{equation}
 \dot{y}_k = \frac{10^{\tau}}{10^{y_k}}(A_k^{ij}n_j+B_k^i) n_i\,.
\end{equation}
and the residuals are naturally defined as
\begin{equation}
\mathcal{R}_{y_k} = \dot{y}_k -\frac{10^{\tau}}{10^{y_k}}(A_k^{ij}n_j+B_k^i)n_i\,. \label{eq:y_residual}
\end{equation}

The relation between the residuals computed in the logarithmic space and in the linear space is
\begin{equation}
    \label{residual_rel}
    \mathcal{R}_{y_k}=\frac{t}{n_k}\mathcal{R}_{n_k}\,.
\end{equation}

Dividing the residuals by $n_k$, the loss function should be balanced, even when the sum is extended to species with order of magnitude difference between abundances (e.g. H$^-$ vs $H$).

Despite these precautions, the loss function remains complex and consequently the training procedure can present mythologies, e.g. excessive stiffness in the parameters update. To mitigate this problem, the weights $\lambda_i$ are modified during the training, using the procedure detailed in \citet{2021SJSC...43A3055W}.
Summarizing, we write the loss function as
\begin{equation}
    \label{eq:loss_composition}
    \mathcal{L}(\theta)=\sum_{i=1}^{N}\lambda_i\mathcal{L}_i(\theta)\,,
\end{equation}
where the sum is extended to $N=2(N_{spec}+1)$ to account for the ODE residuals for the chemical species and the temperature (eq. \ref{eq:loss_f}) and initial conditions (eq. \ref{eq:loss_g}). The terms $\lambda_i$ are adaptively regulated utilizing the back-propagated gradient statistics during model training. 
By experimenting different approaches, we noticed that it is convenient to adopt such adaptive regulation of the weights of the loss, mainly because it improves the convergence during the initial phases of training; this is later discussed in Sec. \ref{sec:results:training}.

\subsection{Network architecture and training strategy}\label{sec:network_struct_ISM}

\begin{figure*}
    \centering
    \includegraphics[width=1\textwidth]{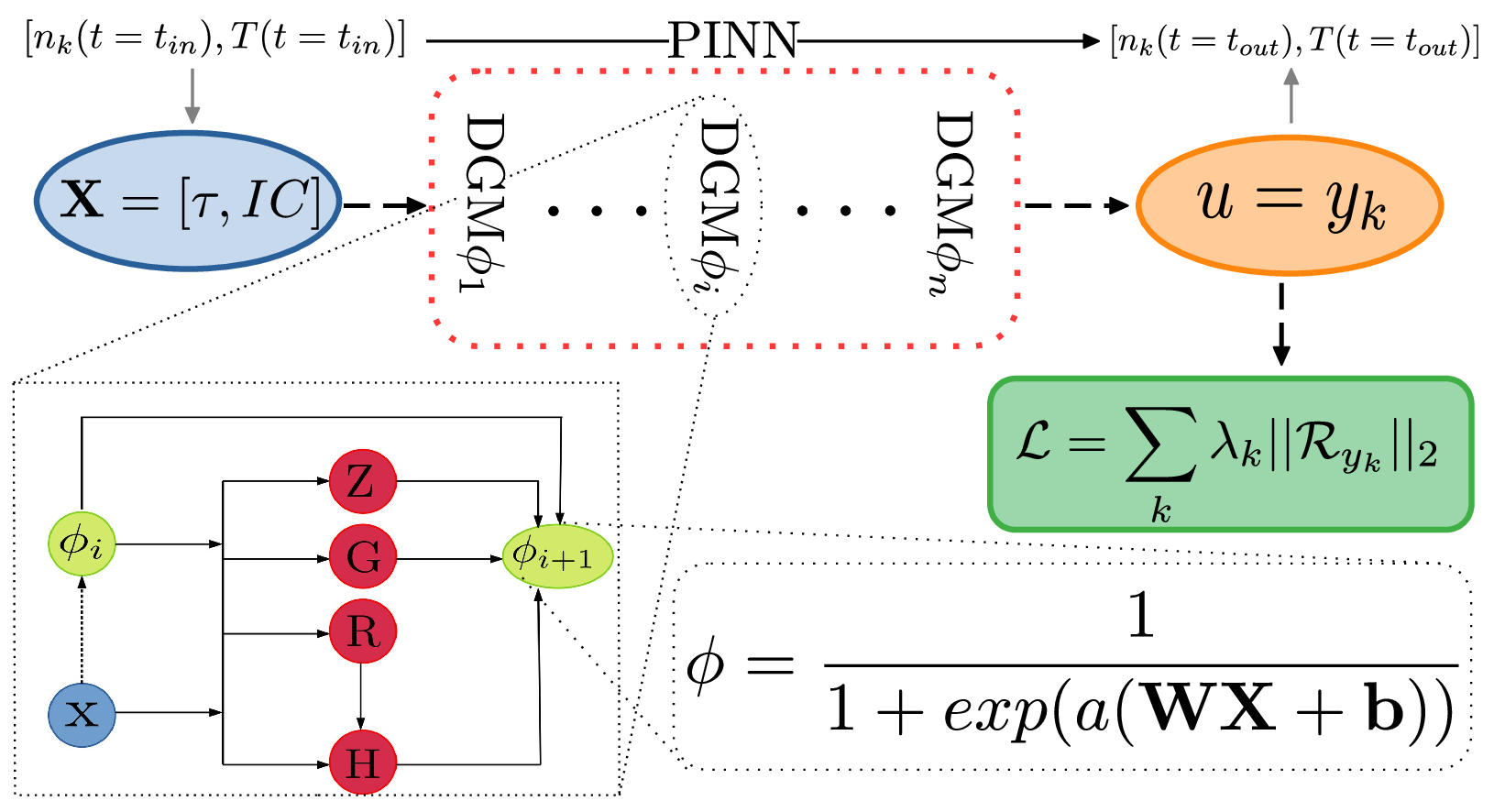}
    \caption{Representation of the PINN model to solve ISM chemistry.
    On the surface, the PINN takes the densities ($n_k$) and temperature ($T$) at initial time $t_{in}$ and returns the evolved quantities at time $t_{out}$ (eq. \ref{eq:map_ISM}); the NN architecture is detailed as follows.
    \textbf{Top}: the model inputs $\mathbf{X}$ (logarithmic time $\tau$ and logarithmic initial conditions $IC$, eq.s \ref{logsystem}) pass through the network layers (${\rm DMG}\phi_i$) and gives the outputs $u$ (the logarithm of abundances and temperature, $y_k$), which is trained to minimize the loss function ($\mathcal{L}$), which is as a linear combination of weighed (eq. \ref{eq:loss_composition}) residuals (eq. \ref{eq:y_residual}).
    \textbf{Bottom left}: inset representing the Deep Galerkin Network (DGM) layer (eq.s \ref{eq:DGMlayer}); $\mathrm{X}$ represents the input data that enters the first layer $\phi_i$ if $i=1$ (dashed line, eq. \ref{eq:first_layer}).
    \textbf{Bottom right}: inset showing the action of the dense layer $\phi$, that is designed using an adaptive sigmoid function (eq. \ref{eq:adaptive_activation}), which depends on the weights $\mathbf{W}$, the biases $\mathbf{b}$, and the adaptive hyperparameter $a$.
    }
    \label{fig:dgm}
\end{figure*}

In general, the optimization of neural network topology and the hyperparameters tuning is a complex task. Although there are tools designed for hyperparameters optimization, an accurate calibration of these is usually not practical, especially if the training procedure is time expensive. So we perform an optimization based on small variations of a fiduciary setting.

For our models, we adopt either a simple Feed-forward Neural Network (FNN, Fig. \ref{PINN_scheme}) or a more advanced DGM architecture \citep{2018JCoPh.375.1339S}.
A DGM follows essentially the same 0)-3) steps described in Sec. \ref{sec:pinn_general}, however the parameters update is more complex (see sketch in Fig. \ref{fig:dgm} ). At depth $i^{th}$ first we calculate the output of a standard dense layer
\begin{equation}\label{eq:first_layer}
\phi_i(\mathbf{x})=\sigma(\mathbf{W}_i\mathbf{x}+\mathbf{b}_i)\,.
\end{equation}
The result is then processed through a DGM layer by computing $Z_i, G_i, R_i$, and $H_i$ as follows:
\begin{subequations}\label{eq:DGMlayer}
\begin{align}
    Z_i &=\sigma(\mathbf{U}^{(z)}_i\mathbf{x}+\mathbf{W}^{(z)}\phi_i+\mathbf{b}^{(z)}_i) \\
    G_i &=\sigma(\mathbf{U}^{(g)}_i\mathbf{x}+\mathbf{W}^{(g)}\phi_i+\mathbf{b}^{(g)}_i) \\
    R_i &=\sigma(\mathbf{U}^{(r)}_i\mathbf{x}+\mathbf{W}^{(r)}\phi_i+\mathbf{b}^{(r)}_i) \\
    H_i &=\sigma(\mathbf{U}^{(h)}_i\mathbf{x}+\mathbf{W}^{(h)}(\phi_i\odot R_i)+\mathbf{b}^{(h)}_i)\,,
\end{align}
\end{subequations}
where $U$ and $\mathbf{W}^{(...)}$ are weight matrices, b are biases, and $\odot$ represent the Hadamard (element-wise) product.
For depth $(i+1)^{th}$, the outputs are then combined via
\begin{equation}\label{eq:layer_up}
\phi_{i+1}(\mathbf{x})=(1-G_i)\odot H_i + Z_i\odot \phi_i
\end{equation}
to define the next dense layer.
A DGM layer requires roughly eight times more memory than standard FNN, since there are eight weight matrices per layer. The main advantage is the ability to capture the \textit{sharp turns} of the underlying solution, as argued in \citet{2018JCoPh.375.1339S}.

For the activation function, we have chosen an adaptive version of the sigmoid function, $\sigma(a,x)$:
\begin{equation}
    \sigma(a,x) = \frac{1}{1+e^{-ax}}
    \label{eq:adaptive_activation}
\end{equation}
where $x$ is the input value and $a$ is a NN adaptive parameter, which is an additional parameter that is optimized during the training. As shown in \citet{JAGTAP2020109136}, this is a useful strategy for dynamical problems that present a wide range of time scales, thus particularly in our case where the reaction rate that can vary by several order of magnitude (see Fig. \ref{fig:matrix_elements}).
The overall architecture design is summarized in Fig. \ref{fig:dgm}.

To fully exploit our hardware, the learning procedure is distributed on multi-GPUs (up to 4) and the train domain is divided in several mini-batches (4-32 per GPU). The optimizer is ADAM, with initial learning rate, $\eta$ and decaying scheduling dependent on the specific model \footnote{Note that the learning rate $\eta$ can be tuned manually on-the-fly for stability reasons.}, which needs a gradient aggregation correction for the parameters update. Furthermore, the initial learning rate, $\eta$, is subjected to a gradual warm-up, following \citet{goyal2018accurate}, which makes $\eta$ less dependent on the user initialization.

The final aspect to address is setup for the training points.
The higher the number of training point, the better the variability of the solution is captured. However, increasing such number does increase the training time and/or the required memory.
It is therefore necessary to find a good trade-off between the amount of training points and computational cost. 
The number is chosen empirically, however, in the most general case of variable initial conditions described in Sec. \ref{sec:initial_conditions}, is shown in \citep{2021arXiv210614473D} that it grows less then exponentially. This result makes us confident that, net of our hardware availability, it is theoretically possible to make eq. \ref{eq:loss_function} converge as the number of chemical species increases.
In this work we find that $\sim 10^4-10^5$ points per GPU for a single batch gives a reasonable balance.
The training points distribution follow the Halton sequence \citep{Halton1964Algorithm2R}; to increase the sampling density, we change the points cloud during the training after a fixed number of iterations, here fixed to 1000 epochs. 
In total we consider 4 cases, which originates from the combination of \textit{single}/\textit{general} models (IC) with \textit{molecular}/\textit{atomic} chemical networks. A summary of the parameters for each model is given in Tab. \ref{tab:models}.
The development of our codes, was perfermod using \code{MODULUS} \citep{hennigh2020nvidia}, a \code{TensorFlow} \citep{tensorflow2015-whitepaper} based tool specific for PINN design.

\section{Results}\label{sec:results}

\begin{figure*}
    \centering
    \includegraphics[width=1\textwidth]{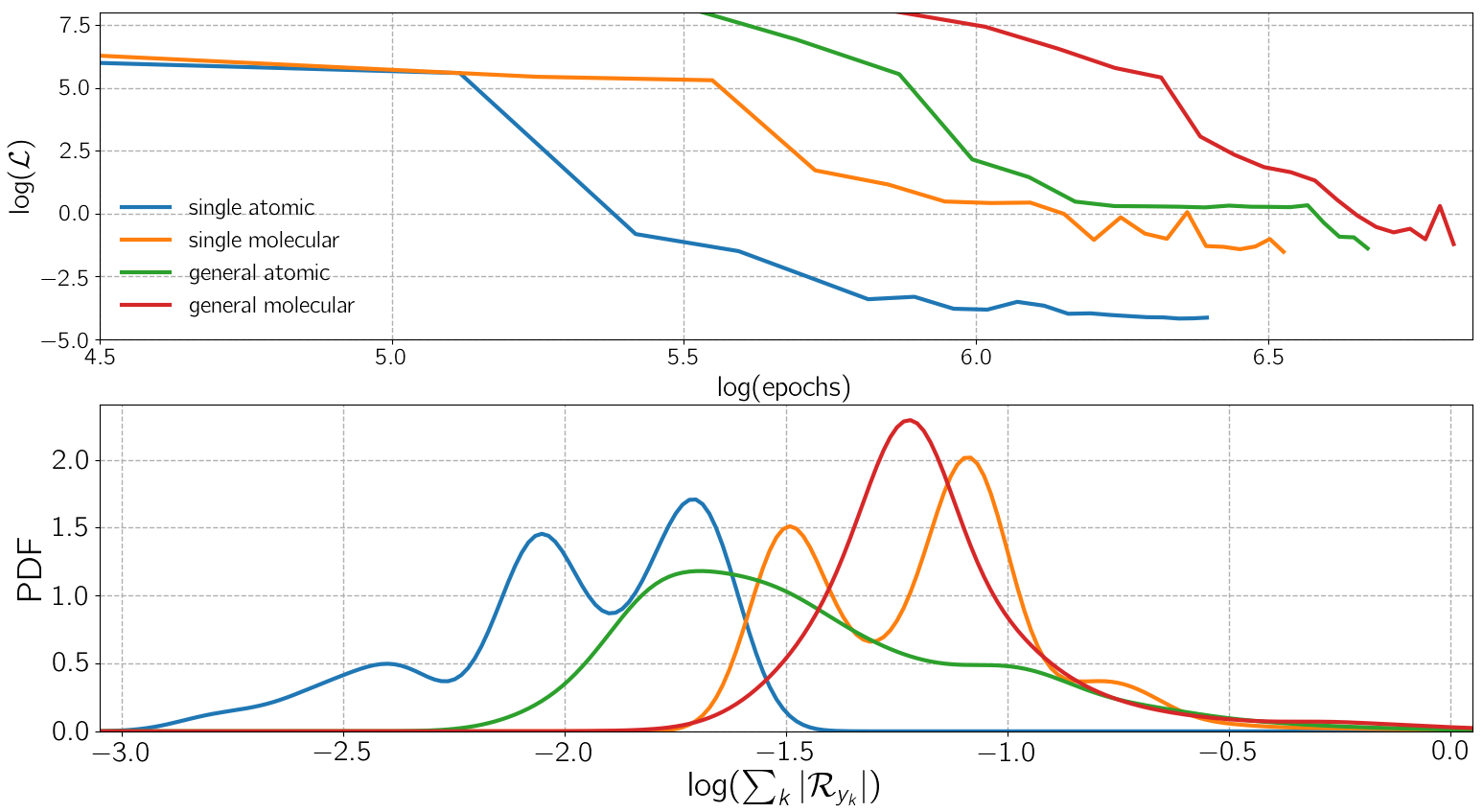}
    \caption{
    Summary of the neural network training.
    \textbf{Top panel}: Loss function ($\mathcal{L}$, eq. \ref{eq:loss_function}) evolution as a function of the training epoch. Each model is indicated with a different colour, according to the caption.
    See Tab. \ref{tab:models} for the main parameters of the models.
    \textbf{Botton panel}: Probability distribution functions (PDF) of sum of the residuals ($\sum_k |\mathcal{R}_k|$, eq. \ref{eq:y_residual}) after the training.
    PDFs are computed by using a kernel density estimation of the absolute value of the logarithm of the residual.
    \label{fig:1Dtrain}
    }
\end{figure*}

\begin{table}
    \centering
    \begin{tabular}{llllll}\\
    \hline
    &   atomic & molecular & atomic & molecular \\
    &   single & single    & general & general \\
    \hline
    chemical system &  ~ &  ~ &  ~ & ~ \\
    \hline
    
    $N_s$ & 7 & 9 & 7 & 9 \\
    $N_{rea}$ & 24 & 46 & 24 & 46 \\
    \textbf{IC} fix & yes & yes & no & no\\
    \hline
    NN architecture &  ~ &  ~ &  ~ & ~ \\
    \hline
    $N_{layers}$ (DGM) & 6 & 8 & 8 & 10\\
    $N_{batch}$  & 16 & 32 & 64 & 128\\
    training points & $1.4\times 10^5$ &  $1.4\times 10^5$ & $2\times 10^5$ & $2\times 10^5$\\
    (per batch) \\
    convergence time &  $120$ & $820$ & $1224$ & $1920$\\  
    (GPUhr) \\
    \hline
    \end{tabular}
    \caption{Reference parameters for the chemical and neural network adopted in the present work.
    The difference between {\it atomic} and {\it molecular} chemical networks is detailed in Sec. \ref{sec:chemical_networks_ISM}. The difference between {\it single} and {\it general} model lays in the type of initial conditions, and is introduced in Sec. \ref{sec:initial_conditions}. The different combinations give a total of 4 models.
    \label{tab:models}
    }
\end{table}

To present our results, first we analyze the training procedure, by studying the convergence of the various model (Sec. \ref{sec:results:training}) and detailing the trends for individual ions (Sec. \ref{sec:results:training_detailed}).
Then, we validate our models against procedural solvers by adopting \code{krome} as our ground truth: first we focus on the \textit{single atomic} network (Sec. \ref{sec:results:validation}), then we give a comparative analysis of the remaining models (Sec. \ref{sec:results:comparison}).

\subsection{Training: convergence}\label{sec:results:training}

An overview of the PINN training can be appreciated in Fig. \ref{fig:1Dtrain}. In the upper panel, we plot $\mathcal{L}$ as a function of the training epochs for all models. 
For all models, the trend of loss functions during the training phase appears qualitatively similar in shape, in particular with a sharp decline in the early epochs.
In general, more complex models (from \textit{single atomic} to \textit{general molecular}) require more training resources to reach an acceptable convergence.
In particular the \textit{single atomic} model is much easier to emulate, in fact the convergence is about two orders of magnitude better with $\sim 5\times10^5$ less training epochs. Furthermore, 

Recall that the training is stopped when $\mathcal{L}$ is below a certain threshold, however there is no generally accepted value for such threshold. Being able to check the residual interactively, it is possible to have an on-the-fly evaluation of the reliability of the approximation.
Additionally, note that if the loss function is stable for a long period ($\sim 10^5 \mathrm{epochs}$), it is possible -- and convenient -- to stop the training and change some hyper-parameters.
In particular, this is an important optimization strategy to adopt in the final stages of the training: when the ADAM algorithm is no longer able to lower the loss function, it is possible to switch to the second order method L-BFGS that uses the hessian matrix; this algorithm is slower but more powerful in the minimization, thus can be used during the last epochs to refine the convergence \citep{bfgs_cite, schraudolph:2007}; we adopt such strategy and report the effect later in Sec. \ref{sec:results:validation}.
further, to quantify the importance of the adaptive weights (see Sec. \ref{sec:loss_ISM}), we perform a control training for the \textit{single molecular} model; without adaptive weights, after $\sim 3.2\times 10^5$ epochs the loss function value is $\sim 20\times$ larger with respect to results shown in Fig. \ref{fig:1Dtrain}.

In the lower panel of Fig. \ref{fig:1Dtrain} we plot the PDF of the sum of the residuals at the last training epoch. In general the peak is around $\sim 10^{-1.5}$  for all models except , with high value tails that can reach up to $\sim 10^{-1}$, with larger residuals for models with higher complexity.
Table \ref{tab:models} summarizes the main hyper-parameters and the convergence time for each model.  Apart from the \textit{single atomic} model, the training time is around thousands of hours (despite the high performance of the GPU used). However, real time can be linearly reduced with the use of multi-GPUs. We also noticed that the \textit{single molecular} and \textit{general atomic} models exhibit similar results in terms of convergence (same number of layers and similar training time). 
While it is a convenient and compact comparison, summing the residual for each chemical species does hide some interesting aspects of the convergence.
\subsection{Species by species convergence}\label{sec:results:training_detailed}

\begin{figure*}
    \centering
    \includegraphics[width=1\textwidth]{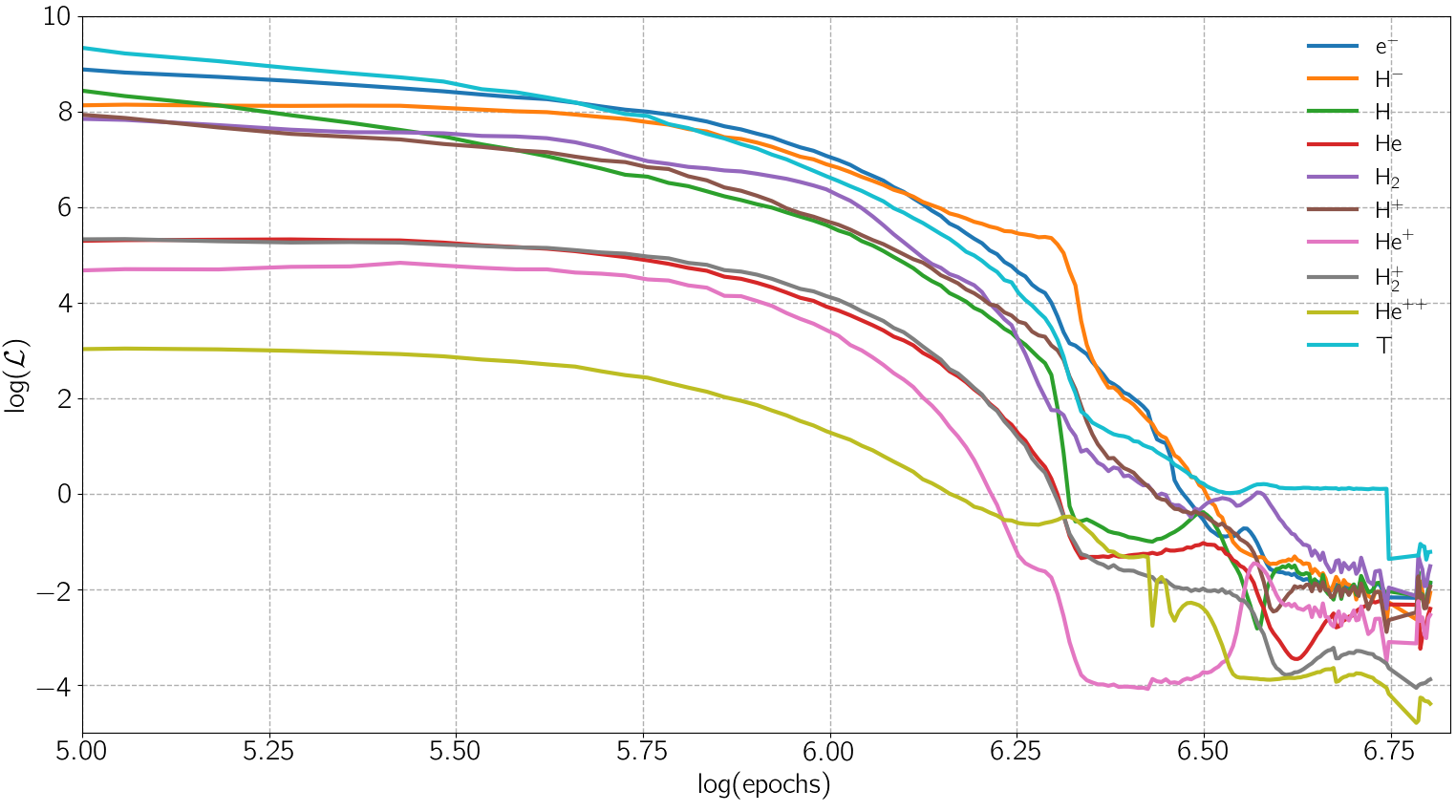}
    \caption{Species by species loss function during the training of the \textit{general molecular} model. Notation is analogue to the one in upper panel of Fig. \ref{fig:1Dtrain}, however the sampling of the loss at different epoch is finer.
    \label{fig:individual loss}
    }
\end{figure*}

\begin{figure*}
    \centering
    \includegraphics[width=1\textwidth]{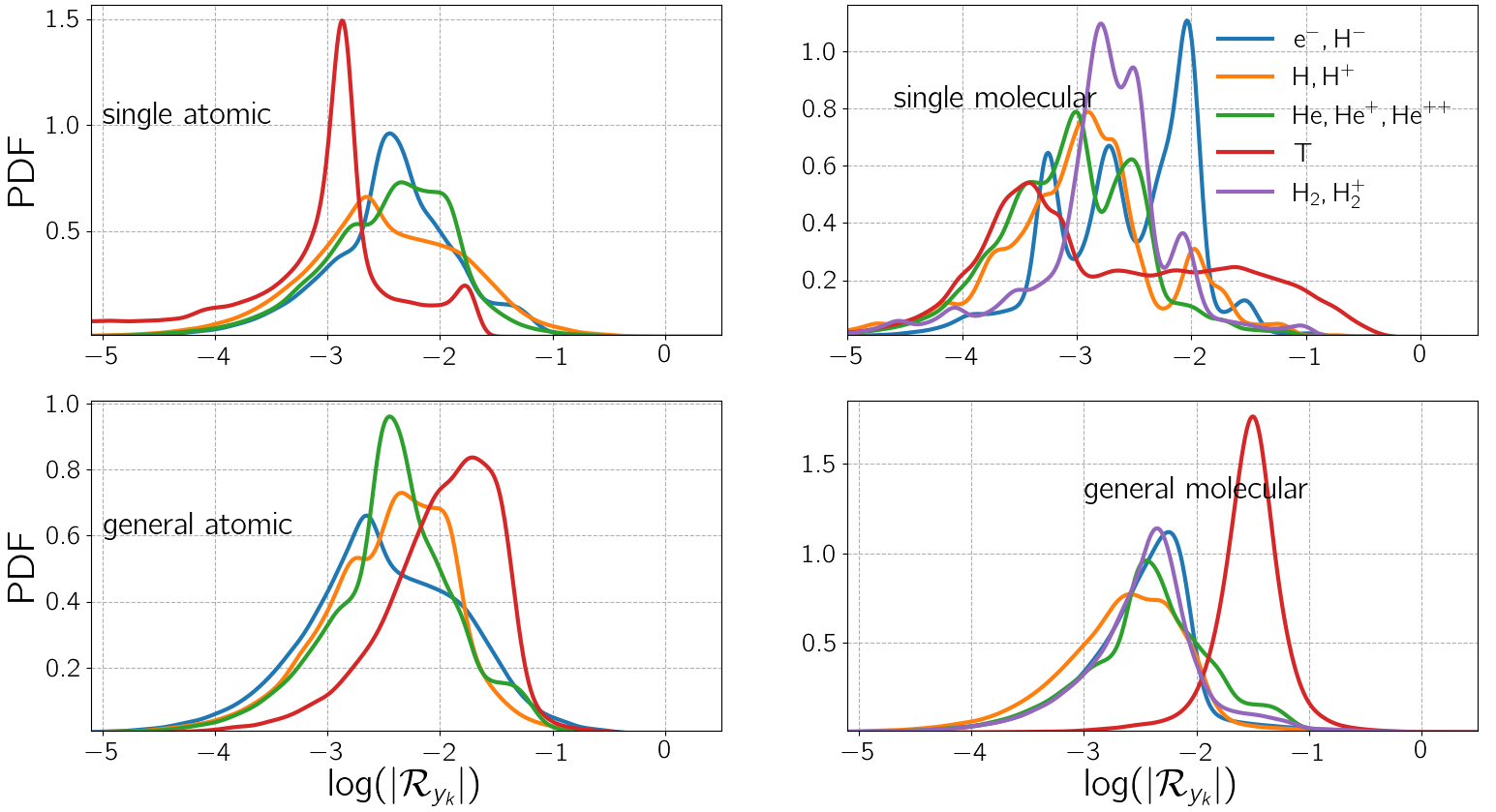}
    \caption{
    Species by species residuals ($\mathcal{R}_k$) after the training. In each panel we show the PDF for a different PINN model, as indicated in the inset. PDF of different contributions to $\mathcal{R}_k$ are grouped with different colours, according to the legend (top right panel). See Tab. \ref{tab:models} for references. 
     \label{fig:KDE_residual}
    }
\end{figure*}

Unlike the most ML applications, a simple evaluation of the loss function (both for training and validation points) it is not enough to guarantee the goodness of the model for PINN algorithm. Indeed, the loss function gives a domain-wide average, thus does not imply local convergence; this aspect is particularly important since we adopt multi-components loss function thus -- despite the precautions taken (Sec. \ref{sec:loss_ISM}) -- some chemical species or a particular set of \textbf{IC} might be not well reconstructed.

Fig. \ref{fig:individual loss} shows an example of the training of each of the components of the loss function in the \textit{general molecular} case. Up to $\simeq 1.2\times 10^6$ training epochs, all the different components follow a similar decreasing trend. After $\simeq 4\times 10^6$ the model start to converge, and the different components saturates at different values; in particular, we can see that the temperature reconstruction dominates the loss function for this specific model. To assess the situation, it is convenient to look at the distribution of residuals when the training is completed.

In the upper panels of Fig. \ref{fig:KDE_residual} we show the various components of the loss function of the for both the {\it atomic} and {\it molecular} single model cases.
For the {\it single atomic} model, all individual residuals are smaller than about $10^{-2}$, with the exception of a negligible contribution for the high values tail of the distributions.
For the \textit{single molecular} model, in general similar values are present, but the high values tails are in general larger, as expected from different values of the loss function when the training is stopped; in particular, the high values tail of temperature evolution surpasses $10^{-2}$.

Recall that, for \textit{general models}, \textbf{ICs} span $20 \leq T_{in}/\mathrm{K} \leq 10^5$ and $10^{-2} \leq n_{in}/\mathrm{cm}^{-3}\leq 10^{3}$, and the fraction of each species is selected to respect charge neutrality and the constraint $X \simeq 0.7$ and $Y\simeq 0.3$, i.e. they have larger number of dimension with respect to \textit{single models}.
Looking on the bottom-left panel of Fig. \ref{fig:KDE_residual}, the residual distributions for all species in the case of \textit{general atomic} model are centered around (or below for negative particles) $10^{-2}$ with tails that does not exceed $10^{-1}$. For \textit{general molecular} model (bottom-right panel of Fig. \ref{fig:KDE_residual}), the residuals of shown a similar behavior except temperatures show similar errors except temperature distribution, centered at $\sim 10^{-0.5}$ (compare with Fig. \ref{fig:individual loss}).
From Fig. \ref{fig:KDE_residual} we can draw the following conclusions: (\textbf{i}) except for \textit{single model} the temperature evolution residuals are the most difficult to minimize and (\textbf{ii}) the residual distribution are for different species are about the same order of magnitude, which implies a good balance of the loss function.

\subsection{Validation of the single atomic network}\label{sec:results:validation}

To validate the training, we compare with the results from \code{KROME} \citep{grassi:2014}, that it is used to build the procedural solvers for both the {\it atomic} and {\it molecular} chemical networks (see Sec. \ref{sec:metodo} for a description of the code). 

In Fig. \ref{fig:comparison1D} we show the evolution of the \textit{single atomic} model for $1 \mathrm{Myr}$.  We note a  qualitative good agreement with procedural solvers in terms of logarithm of the density except for $\mathrm{He}^{++}$ and $\mathrm{H}^-$. We have obtained a very good reconstruction of the initial conditions (which are a soft constraint in the model) and in general the algorithm mixes the variations in density with precision that decreases in cases of species with sharp turns (in log-space).

The choice of adopting logarithmic time (eq. \ref{eq:sub:timelog}) plays an important role here; while it has no immediate advantage in terms of the loss function (eq. \ref{residual_rel}), it helps in the recovery of abundance evolution which starts far from the equilibrium, which consequently have a very steep -- and thus difficult to reproduce -- evolution. In practical terms, the logarithmic time scale has a smoothing effect on these sharp turns and convergence is thus facilitated. However, despite the smoothing effect, the fastest varying species have the largest relative errors.

Note that in the case show in Fig. \ref{fig:comparison1D}, a non-negligible benefit comes from refining the training with a L-BFGS optimizer \citep{bfgs_cite, schraudolph:2007}.
This change lowered the training curve only by a factor $\simeq 1/3$; this is expected: with respect to ADAM, L-BFGS is of higher order but is more prone to get stuck in local minima when the ODEs are stiff \citep{lu2020deepxde}.
Importantly, the adoption of the L-BFGS refinement has significantly improved the validation with \code{KROME}. In particular, before $\mathrm{He^{++}}$ and $\mathrm{H^-}$ showed errors two order of magnitude larger. Further, by using L-BFGS, the reconstruction of the species at early time and at equilibrium is improved.

To be more quantitative, it is convenient to define both relative ($\Delta_r$) and fractional ($\Delta_f$) differences:
\begin{subequations}\label{eq:error_definition}
\begin{align}
    \Delta_f &= \frac{|n_{NN}-n_{KR}|}{n_{tot}} \\
    \Delta_r &= \frac{|n_{NN}-n_{KR}|}{n_{KR}}\,,
\end{align}
\end{subequations}
where $n_{NN}$ are the values predicted with the trained models and $n_{KR}$ are computed with \code{KROME}.
In the bottom panel of Fig. \ref{fig:comparison1D} we show the distribution of relative and fraction errors.
While for most of the species the relative error are reasonably small ($\Delta_r \lsim 10^{-1}$), the relative errors are dominated by $\mathrm{H}^-$, which peaks at $\Delta_r \sim 1$ and $\mathrm{He}^{++}$, which has a very high values tail.
However, these two species have low abundances, thus the fractional errors of all species are small, i.e. $\Delta_f \lsim 10^{-2}$. 

In terms of usage, it seems encouraging that the dominant relative errors affect the less abundant species, since they lead to negligible fractional errors; however, low abundance species can be important in some situations, i.e. H$^-$ abundance is critical in computing H$_2$ in low metallicity/dust environments. 
For a proper usage, such errors should be removed with further training and/or by reconsidering the architecture via a change of hyperparameters setup.

\begin{figure*}
    \centering
    \includegraphics[width=1.0\textwidth]{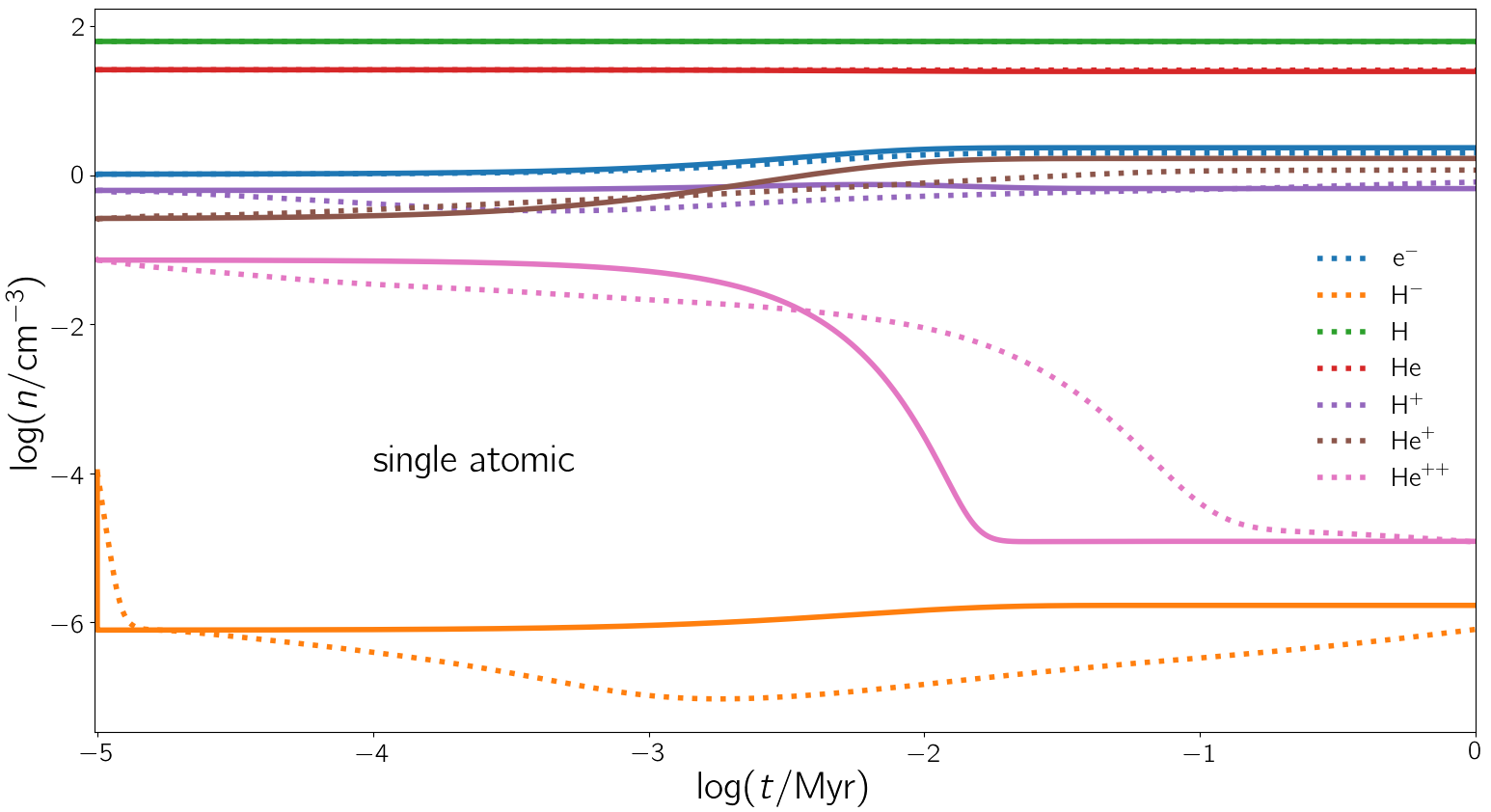}
    \includegraphics[width=0.99\textwidth]{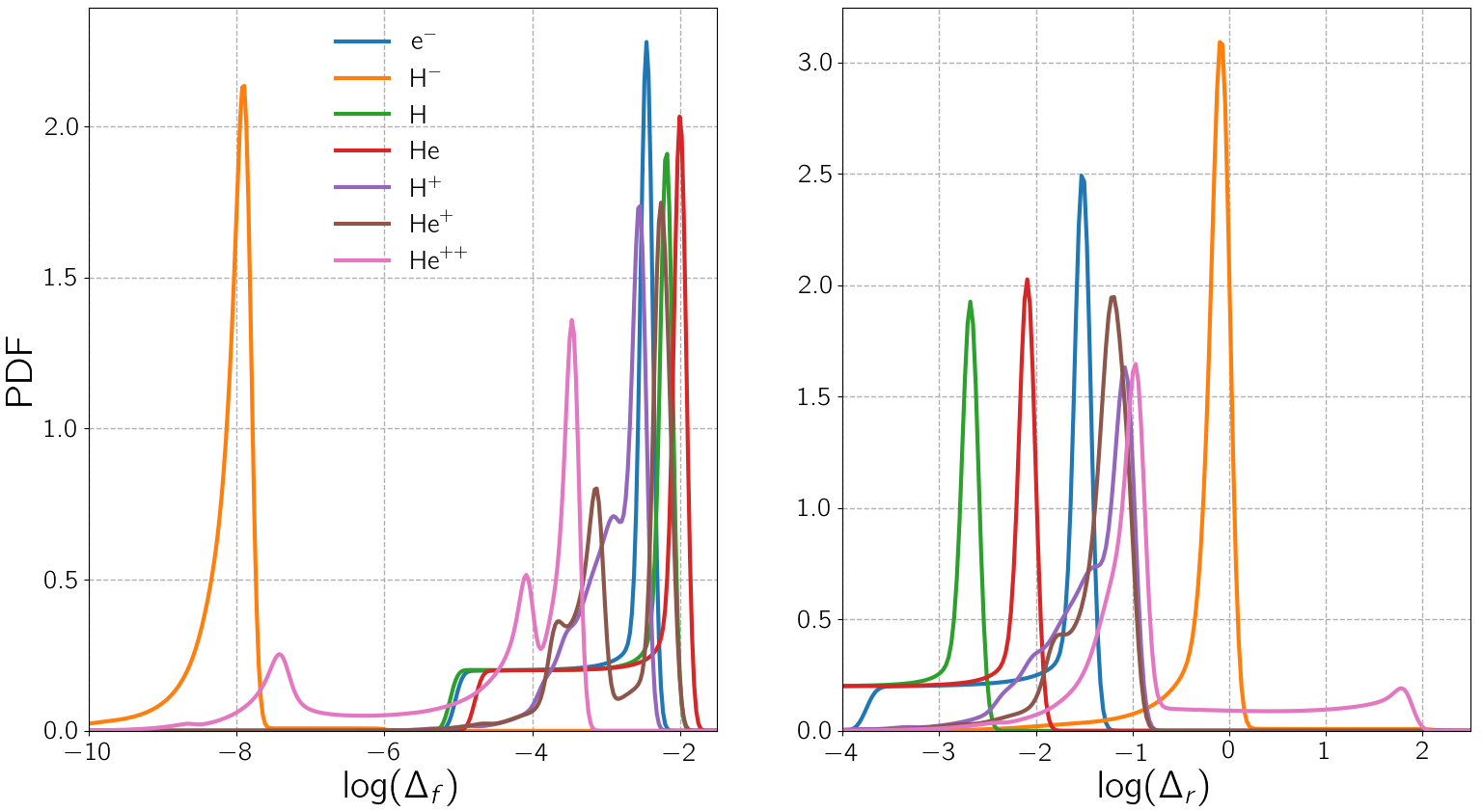}
    \caption{Model comparison between our PINN and \code{KROME} for the \textit{singol atomic} model.
    \textbf{Upper panel}: number density ($n_k$) as a function of time ($t$) for each species. Solid lines represent the solution from \code{KROME}, while the dotted lines shown the PINN predictions. 
    \textbf{Bottom panels}: PDF of the differences between the PINN and \code{krome}. In the left (right) panel, we shows the absolute (relative) $\Delta_f$ ($\Delta_r$) difference for each species. For the definitions, eq.s \ref{eq:error_definition}.
    \label{fig:comparison1D}    
    }
\end{figure*}

\subsection{Model comparison}\label{sec:results:comparison}

\begin{figure*}
  \centering
  \includegraphics[width=1.0\textwidth]{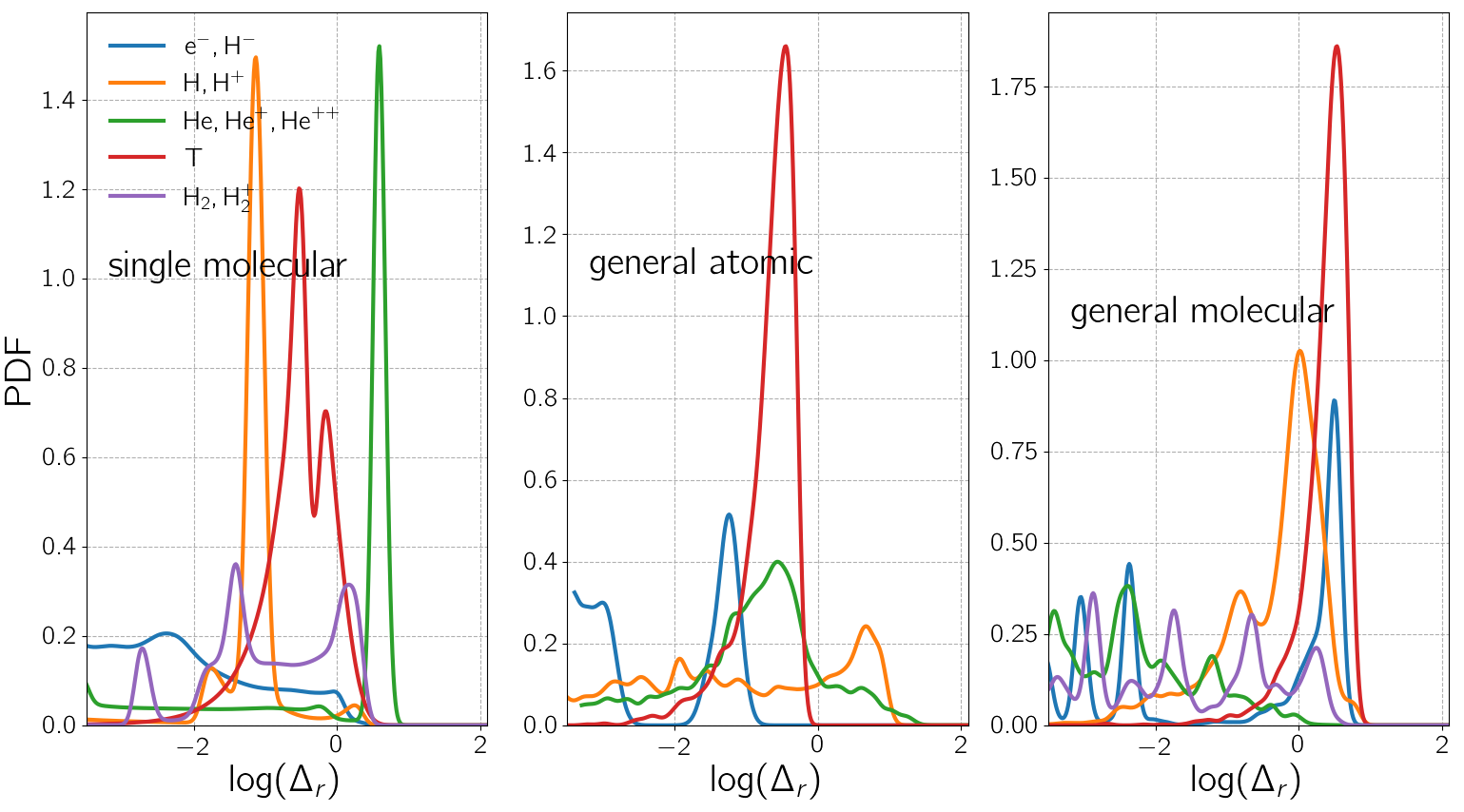}
  \caption{Relative errors distribution between \code{KROME} and the following trained PINN models: \textit{single molecular} (left panel), \textit{general atomic} (central panel) and \textit{general molecular} (right panel). As done Fig. \ref{fig:comparison1D}, different species grouped for clarity.
  \label{fig:error_mutiplot} 
  }
\end{figure*}

\begin{table}
    \centering
    \begin{tabular}{llllll}\\
    \hline
    Quantile &  atomic & molecular & atomic & molecular \\
    &   single & single    & general & general \\
    \hline
    $50\%$ & -1.51 & -2.91 & -3.87 & -1.5 \\
    \hline
     $75\%$ &  -1.05 &  -0.84 & -0.89 & -0.15  \\
    \hline
    $90\%$  &  -0.77 & 0.16 & -0.15 &  0.38\\
    \hline
    \end{tabular}
    \caption{Quantiles of stacked distributions shown in Fig. \ref{fig:error_mutiplot}. We utilize quantiles instead of mean and standard deviation because of the non-trivial shape of the errors distribution. The quantile is defined as in  \citet{article}. The numbers are the logarithm of the relative error value that corresponds to the $x\%$ element of the sorted entire ensemble. 
    \label{tab:quantiles}
    }
\end{table}

\begin{figure*}
    \centering
    \includegraphics[width=\textwidth]{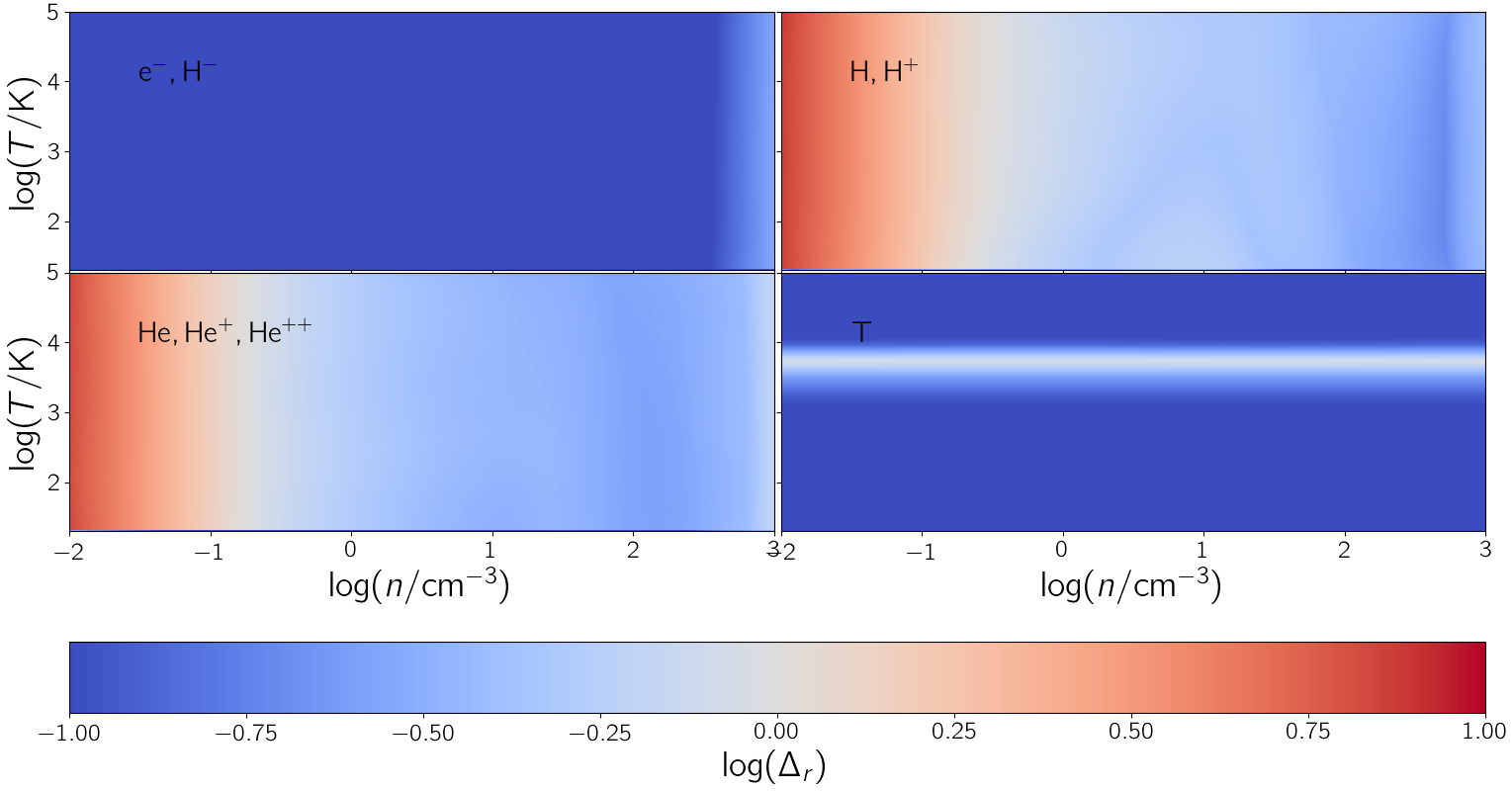}
    \caption{Relative errors for all species in temperature-density ($T-n$) plane for \textit{general atomic} model. Results are grouped in different panels and computed after $\sim 1\mathrm{Myr}$ of evolution from the initial condition. To help the visualization $\Delta_r$ has been cut at the lower (upper) $\min{\log \Delta_r}=-1$ ($\max \log \Delta_r=1$) bound. \label{fig:T_n_multi}
    }
\end{figure*}

\begin{figure*}
    \centering
    \includegraphics[width=\textwidth]{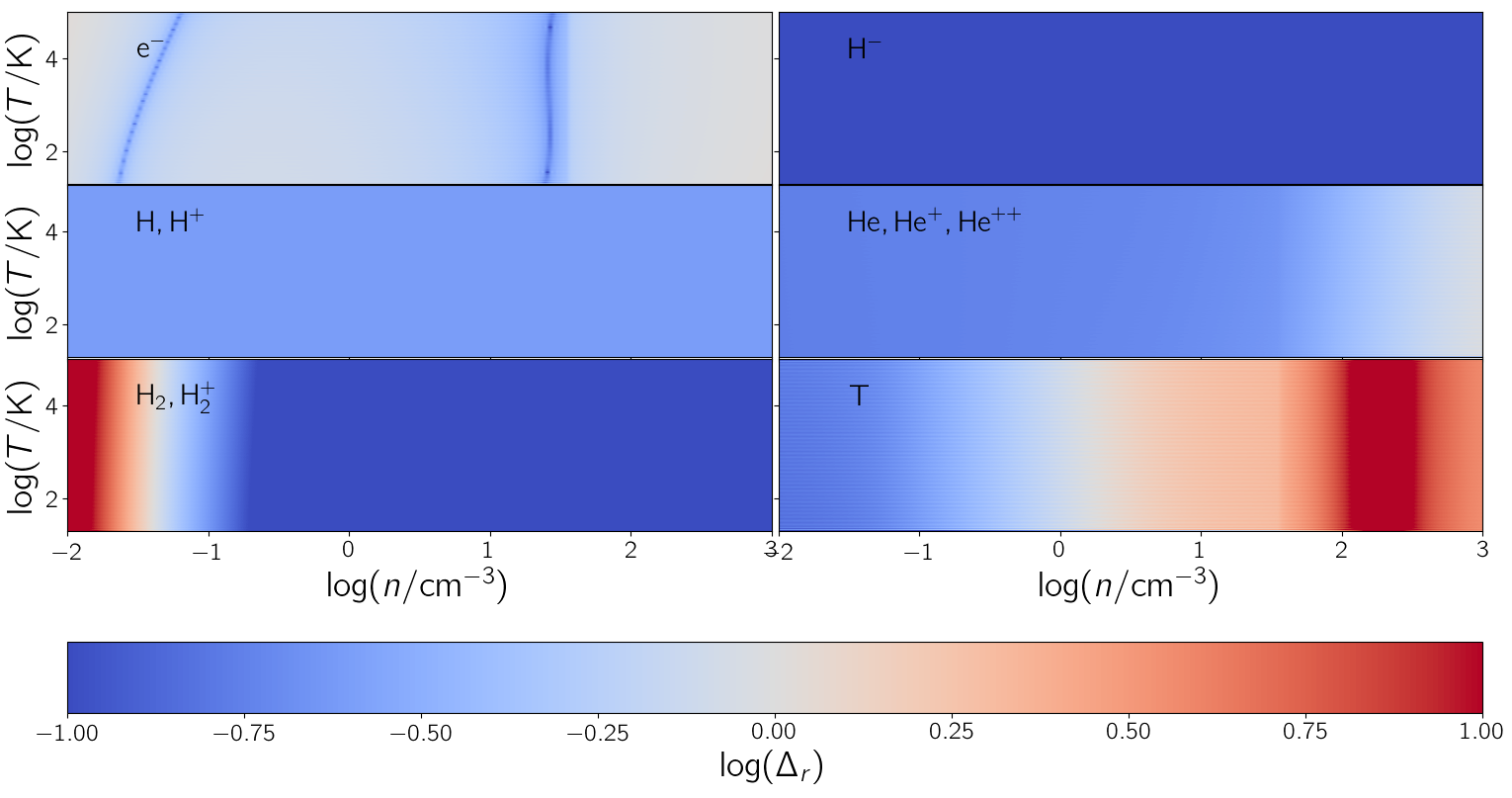}
    \caption{Relative errors for all species in temperature-density ($T-n$) plane for \textit{general molecular} model. Results are grouped in different panels and computed after $\sim 1\mathrm{Myr}$ of evolution from the initial condition. To help the visualization $\Delta_r$ has been cut at the lower (upper) $\min{\log \Delta_r}=-1$ ($\max \log \Delta_r=1$) bound. \label{fig:T_n_multi_mol}
    }
\end{figure*}

Our other models are compared with \code{KROME} in Fig. \ref{fig:error_mutiplot}, where we show the PDF of the relative error $\Delta_r$. For convenience, a summary of the quantiles of the distributions is reported in Tab. \ref{tab:quantiles}.

Results for the \textit{single molecular} model are shown in the left panel of Fig. \ref{fig:error_mutiplot}. Overall we found a good reconstruction, with the $75\%$ of the points with a relative error $\log(\Delta_r)<-0.84$.  However the goodness of the reconstruction presents a species to species variance, in particular we found $\Delta_r\simeq 1$ in the most of cases for $\mathrm{He}^{++}$, similarly to what we found in the \textit{single atomic} model. As this feature is present for both chemical network, it might be a sign that our models fails to accurately capture the ionization of $\mathrm{He}^+$, which has a very small rate given the hardness of the impinging radiation field.

Focusing on \textit{general atomic} model (central panel of Fig. \ref{fig:error_mutiplot}), we note that overall the $75\%$ of the predictions have relative errors smaller than $\log(\Delta_r)<-0.89$; this behaviour is similar to the \textit{single molecular} model, which have been trained for about the same number of epochs.
While \textit{general atomic} has a wider range of {\textbf IC}, \textit{single molecular} have more reactions; however, with respect to other techniques ML is less affected by the curse of dimensionality, thus it is not straightforward to predict a hierarchy of complexity between different models, i.e. larger reactions set vs larger IC parameter space.

In the right panel of Fig. \ref{fig:error_mutiplot} we show the relative errors distribution for the \textit{general molecular} model. We note an overall greater difficulty to emulate the procedural solvers, with the $75\%$ of prediction is smaller than $\log(\Delta_r)<0.05$, with larger errors on temperature, negatively charged ions and hydrogen.
For most of the species, errors for \textit{general molecular} are higher with respect to previously seen models, as a consequence of the increase of complexity for both the number of reactions and parameter space for IC; however He$^{++}$ presents very small errors, thus overall the PINN convergence rate of individual species seems to be not to be directly linked to the reaction rate of the chemical system.

While up to this point we have considered only 1D PDF, it is interesting to see if the emulation power of \textit{general} models is different depending on the initial position in the $T-n$ plane (Figs. \ref{fig:T_n_multi} and \ref{fig:T_n_multi_mol}).
 Compared to \textit{atomic} models we have a better reconstruction of the evolution of $\mathrm{H}$ and $\mathrm{H}^+$ in the \textit{molecular} models. Although we do not have a definitive explanation for this improvement, we are led to think that the inclusion of $\mathrm{H}_2$ and $\mathrm{H}_2^+$ has some black-box effect that allows the model to better reconstruct the evolution of hydrogen.

In Fig. \ref{fig:T_n_multi} we show the $\Delta_r$ distribution of \textit{general atomic} computed at $\sim 1 \mathrm{Myr}$ as a function of the temperature and gas density.
In Fig. \ref{fig:T_n_multi}, we note a good reconstruction over the entire parameter space for negative ions, while for what concerns hydrogen ($\mathrm{H}$ and $\mathrm{H}^+$) and helium ($\mathrm{He}$, $\mathrm{He}^+$ and $\mathrm{He}^{++}$) we have a good agreement at high densities which gradually worsens at low densities.
While this is not shown in the figure, the error at low density is dominated by positively charged ions for both hydrogen and helium.
Regarding the evolution of the temperature, we generally have a very good reconstruction, except when the initial temperature is $T_{in}\simeq 6\times 10^{3}\mathrm{K}$. 

Similarly to \textit{general atomic}, in Fig. \ref{fig:T_n_multi_mol} we show the relative errors to the \textit{general molecular} model

Relative to \textit{general atomic}, the errors for \textit{general molecular} are larger (as also expected from Fig. \ref{fig:error_mutiplot}), but have a very different behaviour in the $n$-$T$ plane.
Both the H, H$^+$ and He, He$^+$, He$^++$ groups have flat distributions of errors, and show relatively a good agreement with \code{krome}.
Electrons have in general higher errors, with two $n$-$T$ narrow stripes of lower errors, likely caused by a higher concentration of training points.
The molecular hydrogen has high relative errors in the low density regime, where its density is negligible, and the temperature shows a tension that gradually rises with density.
Overall, there seem to be no causal connection between regions with large errors and the underlying chemical system.

Summarizing, despite the inherent difficulty of the problem (especially in \textit{general} cases) for a proof of concept work these are encouraging results, as regions where errors are high can be cured with a longer training.
In view of a coupling with numerical simulations and to ease the convergence rate of the training, an exploration the hyperparameters space of the network architecture is needed, along with an adaptive addition of training points in the regions of parameter space where errors are higher.

It is interesting to conclude our analysis by comparing the timing of the PINN network compared to procedural solvers.
For these tests both codes run on similar machines, specifically PINN runs using a single INTEL XEON Gold 6240 CPU with a frequency of $2.60\mathrm{GHz}$, the procedural solver \code{krome} adopts a single INTEL i7-9700 CPU with a frequency $3.00 \mathrm{GHz}$.
For \textit{single} models, we checked the evolution of system using $10^5$ different end time $t_{end}$ from $10\mathrm{yr}$ to $1\mathrm{Myr}$.
With \code{krome}, we find that \textit{single atomic} (\textit{single molecular}) models have a completion time\footnote{Note that a non-negligible amount of time is spent in the warm-up of the solver.} of from $0.05\mathrm{s}$ ($0.09\mathrm{s}$) to $10.56\mathrm{s}$ ($11.16\mathrm{s}$) with increasing $t_{end}$.
With the PINN, we obtain a speed up of a factor $\sim 207$ ($\sim 116$) for \textit{single atomic} (\textit{single molecular}).
For the general models we prepare a 3D grid that is uniformly sampled in log-spaced: the grid features different initial species fractions, $C_k\in [10^{-6},1)$  (100 points), $T_{in}\in[20,10^6]\mathrm{K}$ (512 points), and $n_{in}\in[10^{-2},10^3]\mathrm{cm}^{-3}$ (512 points); different initial conditions are evolved up to $t_{\rm end}=1 \mathrm{Myr}$.
With respect to the \code{KROME} solver, for the PINN we have a speed-up of a factor of $\sim 108$ and $\sim 91$ for the \textit{general atomic} and \textit{general molecular} respectively. 
The speed-up is better in cases of the simpler thermo-chemical system. However, with exploration of the hyperparameter space of the PINN, it should be possible to obtain lighter networks, i.e. needing fewer algebraic operations to compute the output, to achieve a further the speed-up for complex chemical networks.

It is important to note that varying the initial conditions, the \textit{general atomic} (\textit{general molecular}) \code{KROME} yields a model to model standard deviation of $\sim 24\%$ ($\sim 30\%$) with respect to the mean completion time.
For the PINN, the variance is negligible, i.e. less than $10^{-5}\%$. If exploited, this is a critical advantage over procedural solvers, since the latter are prone to yield load balancing problems, while the usage of the PINN can improve the scaling of parallel numerical codes with hydrodynamic coupled with chemistry.

Finally, we compare with other works have tried to adopt machine learning techniques to solve the chemical evolution, i.e. \code{LATENT\_ODE} \citep{grassi2021reducing} and \code{CHEMULATOR} \citep{2021A&A...653A..76H}.
With respect to the chemical network adopted here, these models emulate more complex chemical networks in a higher density range ($\log n/{\rm cm}^{-3}\gsim 1$), i.e. with 33 and 29 species for \code{CHEMULATOR} and \code{LATENT\_ODE}, respectively, even tough \code{LATENT\_ODE} have no temperature dependence nor evolution.
Differently to these methods, our PINN is completely unsupervised, i.e. the PINN does not requires a pre-computed dataset for the solutions (\code{CHEMULATOR}) or a procedural solver in latent space (\code{LATENT\_ODE}).
Further, while both the present and \code{LATENT\_ODE} are developed with hydrodynamic code coupling in mind, \code{CHEMULATOR} is tough to be a faster alternative to photoionization models \citep{rollig:2007}, i.e. its structure is too computationally demanding to be included in numerical simulations \citep{2021A&A...653A..76H}.

Overall our PINN method has better validation errors with respect to \code{LATENT\_ODE} and comparable to \code{CHEMULATOR}.
For the speed-up, the PINN yields performances that are slightly above the one of \code{LATENT\_ODE} ($\times 65$), with both ML techniques being tested against procedural ODE solvers.
The speed-up for \code{CHEMULATOR} is much better ($\times 50000$), but it is obtained against the photoionization code \code{UCLCHEM} \citep{holdship:2017}, which is a much more complex program with respect to an ODE solver.
Final caveat, these comparisons should be taken with care, as all these models are still at a proof of concept stage, thus further optimizations are possible.

\section{Conclusion}\label{conclusions}

Chemical processes are key in regulating the evolution of the interstellar (ISM) and intergalactic medium.
However, in cosmological and astrophysical simulations, finding solutions for thermo-chemistry networks is numerically costly: the systems are stiff and there are orders of magnitude of differences between the time scale of chemical reactions and the ones of astrophysical processes (e.g. gravity and fluid dynamics).
This requires the usage of robust, high-order, multi step backward integrator for ordinary differential equations (ODE), potentially leading to computational bottlenecks for the numerical simulations.

In this work, we explored the possibility to use trained unsupervised physics informed neural networks (PINN) as an alternative to solve or ameliorate such problems.
The main idea consists in expressing the underlying ODEs solution with a neural network (NN) and treat the problem in a variational way, i.e. by minimizing the residuals of the chemical system. This procedure (Sec. \ref{sec:metodo}) has been first introduced in \citet{2019JCoPh.378..686R} but has never been adopted in the context of thermo-chemistry for astrophysical problems.

We first tested the method using benchmark cases (Sec. \ref{sec:benchmark}), then we developed the necessary technical solutions for our case study (Sec. \ref{sec:chemical_networks_ISM}).
We adopted two different thermo-chemical networks that can solve the ISM chemistry with and without molecular hydrogen formation. We build PINN with fixed and arbitrary initial conditions in \textit{single} and \textit{general} models, respectively.
Our main results can be summarized as follows.

\begin{itemize}
    \item A simple feed-forward architecture cannot reconstruct the evolution of a realistic thermo-chemical network. The minimal setup to achieve good results consist in adopting: a Deep Galerkin Method as Neural Network architecture coupled with adaptive sigmoid activation function, adaptive weights in the loss function, an annealing learning rate, and ADAM optimization algorithm (possibly followed by L-BFGS, see Sec.s \ref{sec:loss_ISM} and \ref{sec:network_struct_ISM}). Moreover, a considerable benefit consists in solving the equations in log-space, both for the abundance and time; while the former is a standard normalization technique, a logarithmic time helps for far-from equilibrium situations, as it increase the time sampling.
    \item Even when running on state-of-the-art GPUs, for all models the training time for convergence is relatively large when compared to typical PINN cases of study that are in the literature. We train the models for $\sim 3\times 10^{6}$ epochs, with a total computational time that can vary from $\simeq\,10^2 \mathrm{GPUhr}$ to $\simeq 2\times10^3 \mathrm{GPUhr}$. Unless dedicated resources are allocated, these relative long training times make it expensive to scan the hyperparameter space of the network to improve the convergence and validation.
    \item As expected, the simplest realistic case (\textit{single atomic}) costs much less respect to the most complex case (\textit{general molecular}), both in terms of training time, number of training points and dimension of the network itself (and thus associated memory). However, the hierarchy of complexity between the other model is not clear. On the one hand, the \textit{general atomic} model is multidimensional, thus i) the problem become more stiff as the reaction rates vary more wildly and ii) the curse of dimensions starts to play a role, even tough traditionally ML is less affected with respect to other techniques. On the other hand, in the case of \textit{single molecular} model, there are about double the reactions and molecular heating and cooling processes. It is unclear which of these two factors is dominant in hindering a fast convergence.
    \item We find overall a good agreement with procedural solver. For almost all models, more than 75\% of the points with relative errors less than 15\% and 50\% of the prediction smaller than 1\%; for \textit{single atomic} we have errors smaller than 17\% in 90\% of cases. In the case of \textit{general} models we obtain better results in the high density region of the thermodynamic parameter space.
    \item  For all models, we obtained a significant speed-up respect to the procedural solvers, from $\sim 200$ for the more simple model to $\sim 90$ for the more complex. Furthermore, the very low variance of computational time in different thermodynamic conditions can potentially solve load balancing problems that can occur in the context of massively parallel codes.
\end{itemize}

Although the chemical networks emulated in this work are relatively small (up to 9 ions and  46 reactions), we do not expect to have excessive problems in increasing the number of species, mainly because, with respect to other methods, the PINN suffer less from the curse of dimensionality (i.e. see \citet{2020arXiv200701138M} for the PINN method in up to $\sim 100$ dimensions).
Thud, it would be interesting to test the PINN for possible application to wider chemical networks, i.e. explicit non equilibrium metals evolution following carbon and oxygen chemistry \citep{2010MNRAS.404....2G}: it would entail tracing the evolution of 32 chemical species and 218 reactions; this would represent an interesting possibility for future developments, which should also give a fairer benchmark with respect to other attempts done in emulating chemistry with different ML techniques \citep{grassi2021reducing}.

Limited to our knowledge, the present work is the first case of PINN application for systems of ODEs with this level of complexity.
Thus, this proof-of-concept work is a necessary step before using the trained models as emulators in simulations. In the range of applicability tested here, we can conclude that the models can be used as emulator without a significant loss of precision, as long as further refinement is included for those $n$-$T$ regions where some of the species tracked by the network experience a loss of precision.
For the future, the promising speed-up (up to $\sim 200$) and the absence of variance in the completion time of the calculation make the PINN a very palatable tool for the solution of chemical networks in astrophysical simulations.

\section*{Acknowledgements}
AP acknowledges support from the ERC Advanced Grant INTERSTELLAR H2020/740120.
We gratefully acknowledge computational resources of the Center for High Performance Computing (CHPC) at SNS.
We acknowledge the use of the Python programming language \citep{python3}, Matplotlib \citep{Hunter2007}, NumPy \citep{VanDerWalt2011}, and Scipy \citep{scipy2019}.

\subsection*{Data availability}

The derived data generated in this research will be shared on reasonable requests to the corresponding author.

\bsp	
\label{lastpage}

\bibliographystyle{mnras}
\bibliography{master,codes}     

\begin{thebibliography}{}
\makeatletter
\relax
\def\mn@urlcharsother{\let\do\@makeother \do\$\do\&\do\#\do\^\do\_\do\%\do\~}
\def\mn@doi{\begingroup\mn@urlcharsother \@ifnextchar [ {\mn@doi@}
  {\mn@doi@[]}}
\def\mn@doi@[#1]#2{\def\@tempa{#1}\ifx\@tempa\@empty \href
  {http://dx.doi.org/#2} {doi:#2}\else \href {http://dx.doi.org/#2} {#1}\fi
  \endgroup}
\def\mn@eprint#1#2{\mn@eprint@#1:#2::\@nil}
\def\mn@eprint@arXiv#1{\href {http://arxiv.org/abs/#1} {{\tt arXiv:#1}}}
\def\mn@eprint@dblp#1{\href {http://dblp.uni-trier.de/rec/bibtex/#1.xml}
  {dblp:#1}}
\def\mn@eprint@#1:#2:#3:#4\@nil{\def\@tempa {#1}\def\@tempb {#2}\def\@tempc
  {#3}\ifx \@tempc \@empty \let \@tempc \@tempb \let \@tempb \@tempa \fi \ifx
  \@tempb \@empty \def\@tempb {arXiv}\fi \@ifundefined
  {mn@eprint@\@tempb}{\@tempb:\@tempc}{\expandafter \expandafter \csname
  mn@eprint@\@tempb\endcsname \expandafter{\@tempc}}}

\bibitem[\protect\citeauthoryear{Abadi et~al.,}{Abadi
  et~al.}{2015}]{tensorflow2015-whitepaper}
Abadi M.,  et~al., 2015, {TensorFlow}: Large-Scale Machine Learning on
  Heterogeneous Systems, \url {https://www.tensorflow.org/}

\bibitem[\protect\citeauthoryear{{Asplund}, {Grevesse}, {Sauval}  \&
  {Scott}}{{Asplund} et~al.}{2009}]{2009ARA&A..47..481A}
{Asplund} M.,  {Grevesse} N.,  {Sauval} A.~J.,   {Scott} P.,  2009, \mn@doi
  [\araa] {10.1146/annurev.astro.46.060407.145222}, \href
  {https://ui.adsabs.harvard.edu/abs/2009ARA&A..47..481A} {47, 481}

\bibitem[\protect\citeauthoryear{{Bakes} \& {Tielens}}{{Bakes} \&
  {Tielens}}{1994}]{1994ApJ...427..822B}
{Bakes} E.~L.~O.,  {Tielens} A.~G.~G.~M.,  1994, \mn@doi [\apj]
  {10.1086/174188}, \href
  {https://ui.adsabs.harvard.edu/abs/1994ApJ...427..822B} {427, 822}

\bibitem[\protect\citeauthoryear{Bovino, Grassi, Capelo, Schleicher  \&
  Banerjee}{Bovino et~al.}{2016}]{bovino:2016}
Bovino S.,  Grassi T.,  Capelo P.~R.,  Schleicher D. R.~G.,   Banerjee R.,
  2016, \mn@doi [Astronomy \& Astrophysics] {10.1051/0004-6361/201628158}, A15,
  1

\bibitem[\protect\citeauthoryear{Byrne \& Hindmarsh}{Byrne \&
  Hindmarsh}{1987}]{byrne:19871}
Byrne G.~D.,  Hindmarsh A.~C.,  1987, \mn@doi [Journal of Computational
  Physics] {https://doi.org/10.1016/0021-9991(87)90001-5}, 70, 1

\bibitem[\protect\citeauthoryear{{Cen}}{{Cen}}{1992}]{1992ApJS...78..341C}
{Cen} R.,  1992, \mn@doi [\apjs] {10.1086/191630}, \href
  {https://ui.adsabs.harvard.edu/abs/1992ApJS...78..341C} {78, 341}

\bibitem[\protect\citeauthoryear{{Chantada}, {Landau}, {Protopapas},
  {Sc{\'o}ccola}  \& {Garraffo}}{{Chantada} et~al.}{2022}]{2022arXiv220502945C}
{Chantada} A.~T.,  {Landau} S.~J.,  {Protopapas} P.,  {Sc{\'o}ccola} C.~G.,
  {Garraffo} C.,  2022, arXiv e-prints, \href
  {https://ui.adsabs.harvard.edu/abs/2022arXiv220502945C} {p. arXiv:2205.02945}

\bibitem[\protect\citeauthoryear{{Chardin}, {Uhlrich}, {Aubert}, {Deparis},
  {Gillet}, {Ocvirk}  \& {Lewis}}{{Chardin} et~al.}{2019}]{chardin:2019}
{Chardin} J.,  {Uhlrich} G.,  {Aubert} D.,  {Deparis} N.,  {Gillet} N.,
  {Ocvirk} P.,   {Lewis} J.,  2019, \mn@doi [\mnras] {10.1093/mnras/stz2605},
  \href {https://ui.adsabs.harvard.edu/abs/2019MNRAS.490.1055C} {490, 1055}

\bibitem[\protect\citeauthoryear{{Chen}, {Rubanova}, {Bettencourt}  \&
  {Duvenaud}}{{Chen} et~al.}{2018}]{2018arXiv180607366C}
{Chen} R. T.~Q.,  {Rubanova} Y.,  {Bettencourt} J.,   {Duvenaud} D.,  2018,
  arXiv e-prints, \href {https://ui.adsabs.harvard.edu/abs/2018arXiv180607366C}
  {p. arXiv:1806.07366}

\bibitem[\protect\citeauthoryear{Cybenko}{Cybenko}{1989}]{citeulike:3561150}
Cybenko G.,  1989, \mn@doi [Mathematics of Control, Signals, and Systems
  (MCSS)] {10.1007/BF02551274}, 2, 303

\bibitem[\protect\citeauthoryear{{De Ryck} \& {Mishra}}{{De Ryck} \&
  {Mishra}}{2021}]{2021arXiv210614473D}
{De Ryck} T.,  {Mishra} S.,  2021, arXiv e-prints, \href
  {https://ui.adsabs.harvard.edu/abs/2021arXiv210614473D} {p. arXiv:2106.14473}

\bibitem[\protect\citeauthoryear{{Decataldo}, {Pallottini}, {Ferrara},
  {Vallini}  \& {Gallerani}}{{Decataldo} et~al.}{2019}]{decataldo:2019}
{Decataldo} D.,  {Pallottini} A.,  {Ferrara} A.,  {Vallini} L.,   {Gallerani}
  S.,  2019, \mn@doi [\mnras] {10.1093/mnras/stz1527}, \href
  {https://ui.adsabs.harvard.edu/abs/2019MNRAS.487.3377D} {487, 3377}

\bibitem[\protect\citeauthoryear{{Decataldo}, {Lupi}, {Ferrara}, {Pallottini}
  \& {Fumagalli}}{{Decataldo} et~al.}{2020}]{decataldo:2020}
{Decataldo} D.,  {Lupi} A.,  {Ferrara} A.,  {Pallottini} A.,   {Fumagalli} M.,
  2020, \mn@doi [\mnras] {10.1093/mnras/staa2326}, \href
  {https://ui.adsabs.harvard.edu/abs/2020MNRAS.497.4718D} {497, 4718}

\bibitem[\protect\citeauthoryear{{Draine}}{{Draine}}{1978}]{1978ApJS...36..595D}
{Draine} B.~T.,  1978, \mn@doi [\apjs] {10.1086/190513}, \href
  {https://ui.adsabs.harvard.edu/abs/1978ApJS...36..595D} {36, 595}

\bibitem[\protect\citeauthoryear{{Dropulic}, {Ostdiek}, {Chang}, {Liu}, {Cohen}
   \& {Lisanti}}{{Dropulic} et~al.}{2021}]{dropulic:2021}
{Dropulic} A.,  {Ostdiek} B.,  {Chang} L.~J.,  {Liu} H.,  {Cohen} T.,
  {Lisanti} M.,  2021, \mn@doi [\apjl] {10.3847/2041-8213/ac09ef}, \href
  {https://ui.adsabs.harvard.edu/abs/2021ApJ...915L..14D} {915, L14}

\bibitem[\protect\citeauthoryear{{Flamant}, {Protopapas}  \&
  {Sondak}}{{Flamant} et~al.}{2020}]{2020arXiv200614372F}
{Flamant} C.,  {Protopapas} P.,   {Sondak} D.,  2020, arXiv e-prints, \href
  {https://ui.adsabs.harvard.edu/abs/2020arXiv200614372F} {p. arXiv:2006.14372}

\bibitem[\protect\citeauthoryear{{Galli} \& {Palla}}{{Galli} \&
  {Palla}}{1998}]{1998A&A...335..403G}
{Galli} D.,  {Palla} F.,  1998, \aap, \href
  {https://ui.adsabs.harvard.edu/abs/1998A&A...335..403G} {335, 403}

\bibitem[\protect\citeauthoryear{{Ge}}{{Ge}}{2022}]{2022RAA....22a5004G}
{Ge} J.,  2022, \mn@doi [Research in Astronomy and Astrophysics]
  {10.1088/1674-4527/ac321e}, \href
  {https://ui.adsabs.harvard.edu/abs/2022RAA....22a5004G} {22, 015004}

\bibitem[\protect\citeauthoryear{{Glover} \& {Abel}}{{Glover} \&
  {Abel}}{2008}]{glover:2008}
{Glover} S.~C.~O.,  {Abel} T.,  2008, \mn@doi [\mnras]
  {10.1111/j.1365-2966.2008.13224.x}, \href
  {https://ui.adsabs.harvard.edu/abs/2008MNRAS.388.1627G} {388, 1627}

\bibitem[\protect\citeauthoryear{{Glover}, {Federrath}, {Mac Low}  \&
  {Klessen}}{{Glover} et~al.}{2010}]{2010MNRAS.404....2G}
{Glover} S.~C.~O.,  {Federrath} C.,  {Mac Low} M.~M.,   {Klessen} R.~S.,  2010,
  \mn@doi [\mnras] {10.1111/j.1365-2966.2009.15718.x}, \href
  {https://ui.adsabs.harvard.edu/abs/2010MNRAS.404....2G} {404, 2}

\bibitem[\protect\citeauthoryear{Goyal et~al.,}{Goyal
  et~al.}{2018}]{goyal2018accurate}
Goyal P.,  et~al., 2018, Accurate, Large Minibatch SGD: Training ImageNet in 1
  Hour (\mn@eprint {arXiv} {1706.02677})

\bibitem[\protect\citeauthoryear{{Grassi}, {Bovino}, {Schleicher}, {Prieto},
  {Seifried}, {Simoncini}  \& {Gianturco}}{{Grassi} et~al.}{2014}]{grassi:2014}
{Grassi} T.,  {Bovino} S.,  {Schleicher} D.~R.~G.,  {Prieto} J.,  {Seifried}
  D.,  {Simoncini} E.,   {Gianturco} F.~A.,  2014, \mn@doi [\mnras]
  {10.1093/mnras/stu114}, \href
  {http://adsabs.harvard.edu/abs/2014MNRAS.439.2386G} {439, 2386}

\bibitem[\protect\citeauthoryear{{Grassi}, {Nauman}, {Ramsey}, {Bovino},
  {Picogna}  \& {Ercolano}}{{Grassi} et~al.}{2021}]{grassi2021reducing}
{Grassi} T.,  {Nauman} F.,  {Ramsey} J.~P.,  {Bovino} S.,  {Picogna} G.,
  {Ercolano} B.,  2021, {Reducing the complexity of chemical networks via
  interpretable autoencoders} (\mn@eprint {arXiv} {2104.09516})

\bibitem[\protect\citeauthoryear{{Gunes Baydin}, {Pearlmutter}, {Andreyevich
  Radul}  \& {Siskind}}{{Gunes Baydin} et~al.}{2015}]{baydin2015automatic}
{Gunes Baydin} A.,  {Pearlmutter} B.~A.,  {Andreyevich Radul} A.,   {Siskind}
  J.~M.,  2015, {Automatic differentiation in machine learning: a survey}
  (\mn@eprint {arXiv} {1502.05767})

\bibitem[\protect\citeauthoryear{{Haardt} \& {Madau}}{{Haardt} \&
  {Madau}}{2012}]{2012ApJ...746..125H}
{Haardt} F.,  {Madau} P.,  2012, \mn@doi [\apj] {10.1088/0004-637X/746/2/125},
  \href {https://ui.adsabs.harvard.edu/abs/2012ApJ...746..125H} {746, 125}

\bibitem[\protect\citeauthoryear{Haghighat \& Juanes}{Haghighat \&
  Juanes}{2021}]{Haghighat_2021}
Haghighat E.,  Juanes R.,  2021, \mn@doi [Computer Methods in Applied Mechanics
  and Engineering] {10.1016/j.cma.2020.113552}, 373, 113552

\bibitem[\protect\citeauthoryear{Halton \& Smith}{Halton \&
  Smith}{1964}]{Halton1964Algorithm2R}
Halton J.~H.,  Smith G.~B.,  1964, Commun. ACM, 7, 701

\bibitem[\protect\citeauthoryear{{Hennigh} et~al.,}{{Hennigh}
  et~al.}{2020}]{hennigh2020nvidia}
{Hennigh} O.,  et~al., 2020, {NVIDIA SimNet\^\{TM\}: an AI-accelerated
  multi-physics simulation framework} (\mn@eprint {arXiv} {2012.07938})

\bibitem[\protect\citeauthoryear{{Hindmarsh}}{{Hindmarsh}}{2019}]{odepack}
{Hindmarsh} A.~C.,  2019, {ODEPACK: Ordinary differential equation solver
  library} (\mn@eprint {ascl} {1905.021})

\bibitem[\protect\citeauthoryear{{Hirashita} \& {Ferrara}}{{Hirashita} \&
  {Ferrara}}{2002}]{hirashita:2002}
{Hirashita} H.,  {Ferrara} A.,  2002, \mn@doi [\mnras]
  {10.1046/j.1365-8711.2002.05968.x}, \href
  {https://ui.adsabs.harvard.edu/abs/2002MNRAS.337..921H} {337, 921}

\bibitem[\protect\citeauthoryear{{Holdship}, {Viti}, {Jim{\'e}nez-Serra},
  {Makrymallis}  \& {Priestley}}{{Holdship} et~al.}{2017}]{holdship:2017}
{Holdship} J.,  {Viti} S.,  {Jim{\'e}nez-Serra} I.,  {Makrymallis} A.,
  {Priestley} F.,  2017, \mn@doi [\aj] {10.3847/1538-3881/aa773f}, \href
  {https://ui.adsabs.harvard.edu/abs/2017AJ....154...38H} {154, 38}

\bibitem[\protect\citeauthoryear{{Holdship}, {Viti}, {Haworth}  \&
  {Ilee}}{{Holdship} et~al.}{2021}]{2021A&A...653A..76H}
{Holdship} J.,  {Viti} S.,  {Haworth} T.~J.,   {Ilee} J.~D.,  2021, \mn@doi
  [\aap] {10.1051/0004-6361/202140357}, \href
  {https://ui.adsabs.harvard.edu/abs/2021A&A...653A..76H} {653, A76}

\bibitem[\protect\citeauthoryear{Hornik, Stinchcombe  \& White}{Hornik
  et~al.}{1989}]{HornikEtAl89}
Hornik K.,  Stinchcombe M.,   White H.,  1989, Neural Networks, 2, 359

\bibitem[\protect\citeauthoryear{{Hu}, {Jagtap}, {Karniadakis}  \&
  {Kawaguchi}}{{Hu} et~al.}{2022}]{2022SJSC...44A3158H}
{Hu} Z.,  {Jagtap} A.~D.,  {Karniadakis} G.~E.,   {Kawaguchi} K.,  2022,
  \mn@doi [SIAM Journal on Scientific Computing] {10.1137/21M1447039}, \href
  {https://ui.adsabs.harvard.edu/abs/2022SJSC...44A3158H} {44, A3158}

\bibitem[\protect\citeauthoryear{Hunter}{Hunter}{2007}]{Hunter2007}
Hunter J.~D.,  2007, \mn@doi [Computing in Science Engineering]
  {10.1109/MCSE.2007.55}, 9, 90

\bibitem[\protect\citeauthoryear{Hyndman \& Fan}{Hyndman \&
  Fan}{1996}]{article}
Hyndman R.,  Fan Y.,  1996, \mn@doi [The American Statistician]
  {10.1080/00031305.1996.10473566}, 50, 361

\bibitem[\protect\citeauthoryear{Jagtap, Kawaguchi  \& Karniadakis}{Jagtap
  et~al.}{2020}]{JAGTAP2020109136}
Jagtap A.~D.,  Kawaguchi K.,   Karniadakis G.~E.,  2020, \mn@doi [Journal of
  Computational Physics] {https://doi.org/10.1016/j.jcp.2019.109136}, 404,
  109136

\bibitem[\protect\citeauthoryear{Ji, Qiu, Shi, Pan  \& Deng}{Ji
  et~al.}{2021}]{Ji_2021}
Ji W.,  Qiu W.,  Shi Z.,  Pan S.,   Deng S.,  2021, \mn@doi [The Journal of
  Physical Chemistry A] {10.1021/acs.jpca.1c05102}, 125, 8098

\bibitem[\protect\citeauthoryear{{Jiang} et~al.,}{{Jiang}
  et~al.}{2020}]{2020arXiv200501463J}
{Jiang} C.~M.,  et~al., 2020, arXiv e-prints, \href
  {https://ui.adsabs.harvard.edu/abs/2020arXiv200501463J} {p. arXiv:2005.01463}

\bibitem[\protect\citeauthoryear{{Jura}}{{Jura}}{1975}]{1975ApJ...197..575J}
{Jura} M.,  1975, \mn@doi [\apj] {10.1086/153545}, \href
  {https://ui.adsabs.harvard.edu/abs/1975ApJ...197..575J} {197, 575}

\bibitem[\protect\citeauthoryear{{Kim}, {Kim}  \& {Ostriker}}{{Kim}
  et~al.}{2018}]{kim:2018}
{Kim} J.-G.,  {Kim} W.-T.,   {Ostriker} E.~C.,  2018, \mn@doi [\apj]
  {10.3847/1538-4357/aabe27}, \href
  {https://ui.adsabs.harvard.edu/abs/2018ApJ...859...68K} {859, 68}

\bibitem[\protect\citeauthoryear{{Kingma} \& {Ba}}{{Kingma} \&
  {Ba}}{2014}]{kingma2014adam}
{Kingma} D.~P.,  {Ba} J.,  2014, {Adam: A Method for Stochastic Optimization}
  (\mn@eprint {arXiv} {1412.6980})

\bibitem[\protect\citeauthoryear{{Kumar} \& {Fisher}}{{Kumar} \&
  {Fisher}}{2013}]{2013MNRAS.431..455K}
{Kumar} A.,  {Fisher} R.~T.,  2013, \mn@doi [\mnras] {10.1093/mnras/stt171},
  \href {https://ui.adsabs.harvard.edu/abs/2013MNRAS.431..455K} {431, 455}

\bibitem[\protect\citeauthoryear{LeCun, Bengio  \& Hinton}{LeCun
  et~al.}{2015}]{lecun2015deeplearning}
LeCun Y.,  Bengio Y.,   Hinton G.,  2015, \mn@doi [Nature]
  {10.1038/nature14539}, 521, 436

\bibitem[\protect\citeauthoryear{{Liu} \& {Nocedal}}{{Liu} \&
  {Nocedal}}{1989}]{bfgs_cite}
{Liu} D.~C.,  {Nocedal} J.,  1989, \mn@doi [Mathematical Programming]
  {10.1007/BF01589116}, 45, 503

\bibitem[\protect\citeauthoryear{{Lu}, {Meng}, {Mao}  \& {Karniadakis}}{{Lu}
  et~al.}{2019}]{lu2020deepxde}
{Lu} L.,  {Meng} X.,  {Mao} Z.,   {Karniadakis} G.~E.,  2019, {DeepXDE: A deep
  learning library for solving differential equations} (\mn@eprint {arXiv}
  {1907.04502})

\bibitem[\protect\citeauthoryear{{Lupi}}{{Lupi}}{2019}]{lupi:2019}
{Lupi} A.,  2019, \mn@doi [\mnras] {10.1093/mnras/stz100}, \href
  {https://ui.adsabs.harvard.edu/abs/2019MNRAS.484.1687L} {484, 1687}

\bibitem[\protect\citeauthoryear{{Maio}, {Dolag}, {Ciardi}  \&
  {Tornatore}}{{Maio} et~al.}{2007}]{maio:2007}
{Maio} U.,  {Dolag} K.,  {Ciardi} B.,   {Tornatore} L.,  2007, \mn@doi [\mnras]
  {10.1111/j.1365-2966.2007.12016.x}, \href
  {http://adsabs.harvard.edu/abs/2007MNRAS.379..963M} {379, 963}

\bibitem[\protect\citeauthoryear{{Mishra} \& {Molinaro}}{{Mishra} \&
  {Molinaro}}{2020}]{2020arXiv200701138M}
{Mishra} S.,  {Molinaro} R.,  2020, arXiv e-prints, \href
  {https://ui.adsabs.harvard.edu/abs/2020arXiv200701138M} {p. arXiv:2007.01138}

\bibitem[\protect\citeauthoryear{{Mishra} \& {Molinaro}}{{Mishra} \&
  {Molinaro}}{2021}]{2021JQSRT.27007705M}
{Mishra} S.,  {Molinaro} R.,  2021, \mn@doi [\jqsrt]
  {10.1016/j.jqsrt.2021.107705}, \href
  {https://ui.adsabs.harvard.edu/abs/2021JQSRT.27007705M} {270, 107705}

\bibitem[\protect\citeauthoryear{{Moseley}, {Markham}  \&
  {Nissen-Meyer}}{{Moseley} et~al.}{2020}]{2020arXiv200611894M}
{Moseley} B.,  {Markham} A.,   {Nissen-Meyer} T.,  2020, arXiv e-prints, \href
  {https://ui.adsabs.harvard.edu/abs/2020arXiv200611894M} {p. arXiv:2006.11894}

\bibitem[\protect\citeauthoryear{Nakamura, Shiratori, Nagano  \&
  Shimano}{Nakamura et~al.}{2021}]{inproceedings}
Nakamura Y.,  Shiratori S.,  Nagano H.,   Shimano K.,  2021. ,
  \mn@doi{10.11159/htff21.113}

\bibitem[\protect\citeauthoryear{{Nejad}}{{Nejad}}{2005}]{2005Ap&SS.299....1N}
{Nejad} L. A.~M.,  2005, \mn@doi [\apss] {10.1007/s10509-005-2100-z}, \href
  {https://ui.adsabs.harvard.edu/abs/2005Ap&SS.299....1N} {299, 1}

\bibitem[\protect\citeauthoryear{{Pallottini}, {Ferrara}, {Bovino}, {Vallini},
  {Gallerani}, {Maiolino}  \& {Salvadori}}{{Pallottini}
  et~al.}{2017}]{pallottini:2017_b}
{Pallottini} A.,  {Ferrara} A.,  {Bovino} S.,  {Vallini} L.,  {Gallerani} S.,
  {Maiolino} R.,   {Salvadori} S.,  2017, \mn@doi [\mnras]
  {10.1093/mnras/stx1792}, \href
  {http://adsabs.harvard.edu/abs/2017MNRAS.471.4128P} {471, 4128}

\bibitem[\protect\citeauthoryear{{Pallottini} et~al.,}{{Pallottini}
  et~al.}{2019}]{pallottini:2019}
{Pallottini} A.,  et~al., 2019, \mn@doi [\mnras] {10.1093/mnras/stz1383}, \href
  {https://ui.adsabs.harvard.edu/abs/2019MNRAS.487.1689P} {487, 1689}

\bibitem[\protect\citeauthoryear{{Pallottini} et~al.,}{{Pallottini}
  et~al.}{2022}]{pallottini:2022}
{Pallottini} A.,  et~al., 2022, \mn@doi [\mnras] {10.1093/mnras/stac1281},
  \href {https://ui.adsabs.harvard.edu/abs/2022MNRAS.513.5621P} {513, 5621}

\bibitem[\protect\citeauthoryear{{Prelogovi{\'c}}, {Mesinger}, {Murray},
  {Fiameni}  \& {Gillet}}{{Prelogovi{\'c}} et~al.}{2022}]{pregolovic:2022}
{Prelogovi{\'c}} D.,  {Mesinger} A.,  {Murray} S.,  {Fiameni} G.,   {Gillet}
  N.,  2022, \mn@doi [\mnras] {10.1093/mnras/stab3215}, \href
  {https://ui.adsabs.harvard.edu/abs/2022MNRAS.509.3852P} {509, 3852}

\bibitem[\protect\citeauthoryear{{Rackauckas}, {Innes}, {Ma}, {Bettencourt},
  {White}  \& {Dixit}}{{Rackauckas} et~al.}{2019}]{rackauckas2019diffeqfluxjl}
{Rackauckas} C.,  {Innes} M.,  {Ma} Y.,  {Bettencourt} J.,  {White} L.,
  {Dixit} V.,  2019, {DiffEqFlux.jl - A Julia Library for Neural Differential
  Equations} (\mn@eprint {arXiv} {1902.02376})

\bibitem[\protect\citeauthoryear{{Raissi}, {Perdikaris}  \&
  {Karniadakis}}{{Raissi} et~al.}{2019}]{2019JCoPh.378..686R}
{Raissi} M.,  {Perdikaris} P.,   {Karniadakis} G.~E.,  2019, \mn@doi [Journal
  of Computational Physics] {10.1016/j.jcp.2018.10.045}, \href
  {https://ui.adsabs.harvard.edu/abs/2019JCoPh.378..686R} {378, 686}

\bibitem[\protect\citeauthoryear{{Robbins} \& {Monro}}{{Robbins} \&
  {Monro}}{1951}]{Robbins&Monro:1951}
{Robbins} H.,  {Monro} S.,  1951, \mn@doi [Annals of Mathematical Statistics]
  {10.1214/aoms/1177729586}, 22, 400

\bibitem[\protect\citeauthoryear{{R{\"o}llig} et~al.,}{{R{\"o}llig}
  et~al.}{2007}]{rollig:2007}
{R{\"o}llig} M.,  et~al., 2007, \mn@doi [\aap] {10.1051/0004-6361:20065918},
  \href {https://ui.adsabs.harvard.edu/abs/2007A&A...467..187R} {467, 187}

\bibitem[\protect\citeauthoryear{Schraudolph, Yu  \& Günter}{Schraudolph
  et~al.}{2007}]{schraudolph:2007}
Schraudolph N.~N.,  Yu J.,   Günter S.,  2007, in Meila M.,  Shen X.,  eds,
  Proceedings of Machine Learning Research Vol. 2, Proceedings of the Eleventh
  International Conference on Artificial Intelligence and Statistics. PMLR, San
  Juan, Puerto Rico, pp 436--443, \url
  {https://proceedings.mlr.press/v2/schraudolph07a.html}

\bibitem[\protect\citeauthoryear{{Semenov} et~al.,}{{Semenov}
  et~al.}{2010}]{2010A&A...522A..42S}
{Semenov} D.,  et~al., 2010, \mn@doi [\aap] {10.1051/0004-6361/201015149},
  \href {https://ui.adsabs.harvard.edu/abs/2010A&A...522A..42S} {522, A42}

\bibitem[\protect\citeauthoryear{{Shen}, {Madau}, {Guedes}, {Mayer},
  {Prochaska}  \& {Wadsley}}{{Shen} et~al.}{2013}]{2013ApJ...765...89S}
{Shen} S.,  {Madau} P.,  {Guedes} J.,  {Mayer} L.,  {Prochaska} J.~X.,
  {Wadsley} J.,  2013, \mn@doi [\apj] {10.1088/0004-637X/765/2/89}, \href
  {https://ui.adsabs.harvard.edu/abs/2013ApJ...765...89S} {765, 89}

\bibitem[\protect\citeauthoryear{{Sirignano} \& {Spiliopoulos}}{{Sirignano} \&
  {Spiliopoulos}}{2018}]{2018JCoPh.375.1339S}
{Sirignano} J.,  {Spiliopoulos} K.,  2018, \mn@doi [Journal of Computational
  Physics] {10.1016/j.jcp.2018.08.029}, \href
  {https://ui.adsabs.harvard.edu/abs/2018JCoPh.375.1339S} {375, 1339}

\bibitem[\protect\citeauthoryear{{Sitzmann}, {Martel}, {Bergman}, {Lindell}  \&
  {Wetzstein}}{{Sitzmann} et~al.}{2020}]{2020arXiv200609661S}
{Sitzmann} V.,  {Martel} J. N.~P.,  {Bergman} A.~W.,  {Lindell} D.~B.,
  {Wetzstein} G.,  2020, arXiv e-prints, \href
  {https://ui.adsabs.harvard.edu/abs/2020arXiv200609661S} {p. arXiv:2006.09661}

\bibitem[\protect\citeauthoryear{{Smith} et~al.,}{{Smith}
  et~al.}{2017}]{smith:2017}
{Smith} B.~D.,  et~al., 2017, \mn@doi [\mnras] {10.1093/mnras/stw3291}, \href
  {https://ui.adsabs.harvard.edu/abs/2017MNRAS.466.2217S} {466, 2217}

\bibitem[\protect\citeauthoryear{{Srivastava}, {Greff}  \&
  {Schmidhuber}}{{Srivastava} et~al.}{2015}]{2015arXiv150500387S}
{Srivastava} R.~K.,  {Greff} K.,   {Schmidhuber} J.,  2015, arXiv e-prints,
  \href {https://ui.adsabs.harvard.edu/abs/2015arXiv150500387S} {p.
  arXiv:1505.00387}

\bibitem[\protect\citeauthoryear{{Tancik} et~al.,}{{Tancik}
  et~al.}{2020}]{2020arXiv200610739T}
{Tancik} M.,  et~al., 2020, arXiv e-prints, \href
  {https://ui.adsabs.harvard.edu/abs/2020arXiv200610739T} {p. arXiv:2006.10739}

\bibitem[\protect\citeauthoryear{{Theuns}, {Leonard}, {Efstathiou}, {Pearce}
  \& {Thomas}}{{Theuns} et~al.}{1998}]{theuns:1998}
{Theuns} T.,  {Leonard} A.,  {Efstathiou} G.,  {Pearce} F.~R.,   {Thomas}
  P.~A.,  1998, \mn@doi [\mnras] {10.1046/j.1365-8711.1998.02040.x}, \href
  {http://adsabs.harvard.edu/abs/1998MNRAS.301..478T} {301, 478}

\bibitem[\protect\citeauthoryear{{Ucci}, {Ferrara}, {Pallottini}  \&
  {Gallerani}}{{Ucci} et~al.}{2018}]{ucci:2018}
{Ucci} G.,  {Ferrara} A.,  {Pallottini} A.,   {Gallerani} S.,  2018, \mn@doi
  [\mnras] {10.1093/mnras/sty804}, \href
  {https://ui.adsabs.harvard.edu/abs/2018MNRAS.477.1484U} {477, 1484}

\bibitem[\protect\citeauthoryear{Van~Rossum \& Drake}{Van~Rossum \&
  Drake}{2009}]{python3}
Van~Rossum G.,  Drake F.~L.,  2009, Python 3 Reference Manual.
CreateSpace, Scotts Valley, CA, \url
  {https://dl.acm.org/doi/book/10.5555/1593511}

\bibitem[\protect\citeauthoryear{{Virtanen} et~al.,}{{Virtanen}
  et~al.}{2019}]{scipy2019}
{Virtanen} P.,  et~al., 2019, arXiv e-prints, \href
  {https://ui.adsabs.harvard.edu/abs/2019arXiv190710121V} {p. arXiv:1907.10121}

\bibitem[\protect\citeauthoryear{{Wang} \& {Raj}}{{Wang} \&
  {Raj}}{2017}]{2017arXiv170207800W}
{Wang} H.,  {Raj} B.,  2017, arXiv e-prints, \href
  {https://ui.adsabs.harvard.edu/abs/2017arXiv170207800W} {p. arXiv:1702.07800}

\bibitem[\protect\citeauthoryear{{Wang}, {Teng}  \& {Perdikaris}}{{Wang}
  et~al.}{2021}]{2021SJSC...43A3055W}
{Wang} S.,  {Teng} Y.,   {Perdikaris} P.,  2021, \mn@doi [SIAM Journal on
  Scientific Computing] {10.1137/20M1318043}, \href
  {https://ui.adsabs.harvard.edu/abs/2021SJSC...43A3055W} {43, A3055}

\bibitem[\protect\citeauthoryear{van~der Walt, Colbert  \& Varoquaux}{van~der
  Walt et~al.}{2011}]{VanDerWalt2011}
van~der Walt S.,  Colbert S.~C.,   Varoquaux G.,  2011, \mn@doi [Computing in
  Science Engineering] {10.1109/MCSE.2011.37}, 13, 22

\makeatother
\end{thebibliography}

\end{document}